\documentclass[pre,preprint,showpacs]{revtex4-1}
\usepackage{amsmath}
\usepackage{bbm}
\usepackage{graphicx}

\begin{document}

\def\fin{\hbox{${\vcenter{\vbox{            
   \hrule height 0.4pt\hbox{\vrule width 0.4pt height 6pt
   \kern5pt\vrule width 0.4pt}\hrule height 0.4pt}}}$}}

\title{Biofilm growth on rugose surfaces\footnote{Physical Review E 86, 061914, 2012 }}
\author{D. Rodriguez} 
\affiliation{Departamento de Matem\'atica Aplicada,
Universidad Complutense de Madrid; Madrid, Spain}
\author{B. Einarsson} 
\affiliation{Center for Complex and Nonlinear Science, University
of California at Santa Barbara; California, USA}
\author{A. Carpio} 
\affiliation{Departamento de Matem\'atica Aplicada,
Universidad Complutense de Madrid; Madrid, Spain,}
 \affiliation{School of Engineering and Applied Sciences,
 Harvard University; Cambridge, Massachusetts, USA}
\date{Nov 13, 2012}

\begin{abstract}

{A stochastic model is used to assess the effect of external parameters on the development of submerged biofilms on smooth and rough surfaces. The model includes basic cellular mechanisms, such as division  and spreading, together with an elementary description of the interaction with the surrounding flow and probabilistic rules for EPS matrix generation,  { cell decay and adhesion}. 
Insight on the interplay of competing mechanisms as the flow or the nutrient concentration change is gained. { Erosion and growth processes combined produce biofilm structures moving downstream.}
A rich variety of patterns are generated: { shrinking biofilms, patches,} ripple-like structures traveling downstream, fingers, mounds, streamer-like patterns,  flat layers, porous and dendritic structures.   The observed regimes depend on the carbon source and the type of bacteria.  }
\end{abstract}


\maketitle

\section{Introduction}

Many bacterial species adapt to hostile environments forming aggregates 
called biofilms. Formation of bacterial biofilms on solid surfaces in aqueous 
environments is a collective effect of great importance in medicine and engineering. 
In such environments, bacteria may attach themselves to a solid surface and
to each other by mechanisms that need clarification, produce extracellular polymeric substances (EPS) and change their morphology \cite{development}. The result is the formation of a film on the solid surfaces that contains live bacteria, grows on a wide range of hydrated supports, e.g. tissue, plastic, metal, rocks, etc, and is extremely resistant to external aggressions due to antibiotics, chemicals or flows. Once formed, biofilms may grow, expel cells and film fragments that are carried by the flow and may in turn reattach themselves to other solid surfaces colonizing them \cite{development}. 

Biofilms are responsible for { a large percentage of} infections in humans,
ranging from deadly illnesses, e.g. cystic fibrosis and legionellosis, to life threatening infections originated at artificial joints, pacemakers or catheters  
\cite{NIH, davies, slimy}. Biofilm induced damage causes substantial economic losses, due to biocorrosion of aircraft fuselages or metallic structures, biofouling of ship hulls, efficiency reduction in heat exchange-systems, pressure drop and clogs in water systems, food and water poisoning  \cite{water}, and so on. On the other hand, bioremediation seems to be one of the fastest growing fields nowadays. Some bacterial strains feed on a wide variety of toxic pollutants. 
Such microorganisms may be deliberately released to clean up oil spills or to 
purify underground water in farming land and mines  \cite{slimy}. 
At the same time, the development of synthetic biology is paving the way for 
the use of biofilms as biosensors or bioindicators for monitoring the presence 
of certain chemicals in the environment 
\cite{chien,morin,delorenzo2,delorenzo3,mages,strosnider,swaay}. 
Attempts to engineer devices out of biofilms continue to take place. 
Electrooptical devices have already been created \cite{delorenzo1}.
There are efforts to use biofilms emitting optic signals as microsensors in microdevices. 
To indicate correctly the order of magnitude of any variable on the walls they attach to,  biofilms should be kept thin and homogeneous. Pattern formation should be avoided.
If we are to control biofilms grown on specific parts of a device, such surfaces must have well defined properties. In particular, a defined roughness pattern. Standardized manufacturing of substrates is necessary to ensure similar biofilm quality under analogous external conditions. Surfaces with an unknown microstructure are mechanically processed by means of milling machines before being used as substrates. The resulting roughness, determined by the milling cutters, is typically of the same order of magnitude as the size of the bacteria. Therefore, the effect of such roughness patterns on biofilm growth under  different flow conditions must  be accounted for.

Understanding biofilm development and their response to changes in the environment is essential to controlling them in engineering and medical systems, either to destroy them when unwanted, or to better exploit them when beneficial. General models describing biofilm dynamics can be classified in three groups according to their description of bacteria \cite{bookmodels,practicioner,microscope}: continuum, individual based (IbM) or cellular automata (CA) models.  Continuum models treat the biofilm as a material, typically a gel, polymer or viscous fluid \cite{viscoelastic,cogan,brenner}. They are often two-phase models comprising the fluid containing the nutrients and the biofilm \cite{continuum}.  { When available, thin film approximations may be more effective
\cite{seminara}}.
In IbM models, microbes or clusters of microbes are seen as hard spherical particles or macroscopic objects with mechanical properties that evolve according to reaction-diffusion equations for nutrients and oxygen coupled with descriptions of bacterial growth and spreading of biomass \cite{particle}. 
Hybrid variants include the EPS matrix as an incompressible viscous flow, which embeds the discrete microbial cells \cite{hybridmatrix}.    Hybrid models combine discrete descriptions of the cells with continuous descriptions of other relevant fields \cite{plh,poplawski}. A hybrid model trying to account for the effect of the flow is proposed in \cite{hybridhydro}.
CA models distribute biomass over a cellular grid and allow its cells to change with appropriate probabilities according to a set of simple rules \cite{hermanovic,qca}. Available CA models already include a few bacterial mechanisms and activities in a reasonable way. However, the description of more complex mechanisms, such as microbe attachment to surfaces \cite{adhesion1,adhesion2}, quorum sensing to form biofilms \cite{quorum,ieee},  generation of EPS matrix \cite{matrix1,matrix2,cooperation} and  interaction with the surrounding flow \cite{paureaginosa,3d,detachment} remains unclear.  

The most appropriate model for a particular biofilm should take into account the specific bacteria, the environment in which the biofilm is formed, the parameters that can be fitted to experiments, and the predictions to be made \cite{bookmodels,practicioner,microscope}. 
We have in mind { understanding} experiments of submerged biofilm growth in micropipes, on surfaces whose roughness is of the same order of magnitude as the bacterial size.  Despite the vast amount of literature on biofilm modeling, we are not aware of previous attempts to include such roughness patterns in the models. The size of the bacteria being comparable to the magnitude of the roughness, we have chosen to borrow ideas from cellular automata strategies, conveniently modified to account for surface roughness. 
{ To reduce the computational cost, we focus on a two dimensional reduction, though the three dimensional version of the model is discussed too.}
{ We reproduce qualitatively patterns and trends that remind of those observed  in experiments, and are able to make a few additional predictions. 
However, additional processes like  cell displacement within the biofilm or movement of biofilm blocks induced by external forces have yet to be included in the model for a better understanding of pattern formation processes.}
Experimental measurements of some parameters (unavailable at present) should be required to  attempt quantitative comparisons. 
Fitting model parameters to practical set-ups seems to  be a general problem with models for biofilms due to lack of adequate experimental data  \cite{practicioner,poplawski}.
 
The paper is organized as follows. Section 2 collects basic facts on biofilms and fixes the set-up for the problem, based on experiments of submerged biofilm growth. 
Section 3 presents a two dimensional model for aerobic biofilm growth in micropipes, together with the probabilistic rules for cellular division and spreading, detachment due to the flow, EPS matrix generation, { decay and adhesion}. We propose an asymmetric erosion mechanism adapted to flows moving parallel to the surface, together with simple probabilistic rules accounting for  EPS matrix generation and its influence on biofilm cohesion. { Section 4 extends the model to three dimensions}. { Section 5 focuses initially on 
the evolution of  biofilm seeds already attached to the substratum for  different nutrient concentrations
and increasing ratios  of the shear due flow to the biofilm cohesion.  
For large ratios, flat biofilm seeds are washed out on smooth surfaces in three dimensions, see 
Figure \ref{washout}. As the ratio shear/cohesion is decreased or the nutrient concentration is increased, ripple-like and streamer-like patterns are generated, see Figures \ref{ripple} and \ref{streamer}.
Such ripples travel downstream with the current. Streamer-like patterns may detach when they surpass a certain size. For smaller ratios or larger concentrations,  tower-like structures may develop, see Figure \ref{mound}.  The influence of the model parameters on the patterns is investigated in more detail by means of the two dimensional reduction. 
The evolution of the patterns with some parameters seems to follow qualitative trends experimentally observed, by other groups \cite{paureaginosa,stoodleyfigures,stoodleynutrients,stoodleyripples, stoodleysulfate,stoodleycohesion}  and by ourselves \cite{unpublished}. However,  the effect
of  mechanical processes like displacement of cell aggregates due to external forces, thought to be relevant in the dynamics of real biofilm patterns such as streamers, must yet be accounted for.
Biofilm  evolution depends on the combinations of cell processes activated and a small set of parameters. The simulations provide some insight on the effect of competing mechanisms. 
{ Erosion and growth alone are able to produce biofilm structures moving downstream.}
The regimes observed as the flow varies are influenced by the carbon source,  the type of bacteria and the nutrient concentration.  This may lead to flatter biofilms or enhanced pattern
formation. When the current is strong enough compared to the biofilm cohesion, biofilm seeds seem to evolve into thin flat biofilms for most nutrients, which may break into patches and tend to be washed out on smooth surfaces but may be stabilized by  some choices of surface roughness. Numerical simulations suggest that roughness of  a slightly larger size than the bacterial size may enhance biofilm ability to survive and cover surfaces in a more uniform way, see Figures \ref{fig81}-\ref{fig85} and \ref{beta}. 
Other types of roughness may hinder biofilm growth. Adhesion of floating cells
on uncolonized surfaces may produce patchy configurations at low adhesion rates, that
may become wavy biofilm layers at larger adhesion rates.}
Finally, { Section 6} discusses our conclusions and perspectives for future work.

\section{Biofilm structure and experiments of pattern formation}
\label{background}

Biofilms are a survival strategy developed  to create homeostatic conditions which favor bacterial survival in hostile environments.  The first scientific evidence of biofilm existence is due to Antonie Van Leeuwenhoek in 1680, who detected these structures while studying the dental plaque. In 1978,  William Costerton  isolated bacterial cells in completely different bacterial developments from those suspended  or ``planktonic" cells seen until that year in laboratories. He named them biofilms    
 \cite{costerton0}.  Until then, the same bacterial strain was used during several generations, keeping it in optimal growth conditions. This induces the natural genetic selection of bacteria unable to generate biofilms, which are  a natural mechanism to survive in adverse environments.  

\begin{figure}
\centering
\includegraphics[width=7cm]{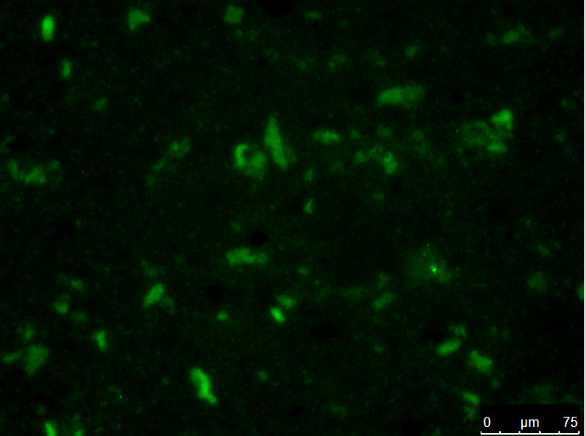}
\caption{Biofilm patches formed by Pseudomonas putida on a polycarbonate
substratum at $Re=100$. Image taken with a confocal LEICA SP5 fluorescence 
microscope.}
\label{fig4}
\end{figure}

Biofilms can be defined as a large amount of microcolonies embedded inside a polysaccharid matrix attached to a solid surface in an aqueous environment. This matrix offers numerous advantages: it gives increased resistance against external aggressions, such as chemicals, high shear flows, radiation, etc., it can be used as a food reservoir, and creates a { suitable} environment to reproduce \cite{liu}. 
{ The external  medium provides the substances bacteria need to live.}
Figure \ref{fig4} shows biofilm patches formed by bacteria genetically modified by inserting the gene for the production of green fluorescent protein (GFP). This protein becomes fluorescent when exposed to light, allowing to locate the biofilm fragments attached to the surface.

\begin{figure}[!t]
\centering
\includegraphics[width=8cm]{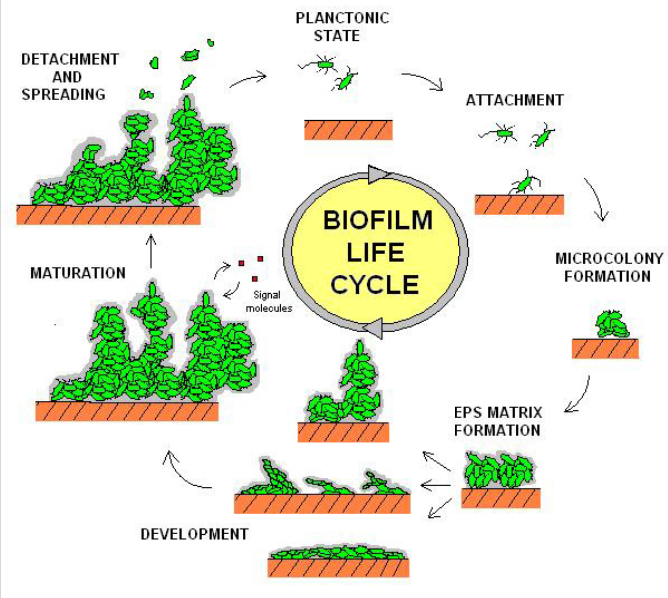}
\caption{Schematic development of a biofilm.}
\label{fig1}
\end{figure}

Since Costerton's work, the field has developed { continuously \cite{development,slimy,matrix1,matrix2,brenner,costerton1,survival,costerton2,kolter,klapper,battin}.} 
However, many issues concerning biofilm behavior and its responses   to changes in the environment remain unresolved. The life cycle of  biofilms has been largely studied \cite{ghannoum, development}. The main developmental stages are: colonization, growth, spread, maturation and death. A scheme of a submerged biofilm life cycle is shown in Figure \ref{fig1}. { Different bacterial species may create different types of biofilms depending on the growth conditions. Biofilms may stick on air/solid or air/liquid interfaces \cite{kolter} forming wrinkled films. We consider here submerged biofilms attached to solid surfaces surrounded by a flow, formed by aerobic bacteria.}

The colonization of new surfaces is a critical stage of the biofilm development. Bacteria have the ability to detect concentration gradients (which are related to the presence of a surface) \cite{costerton1,adhesion2}. Microorganisms use cilia or flagella to reach the surface, trying to attach to it.  The attachment process induces several changes in the morphology of bacteria and their behavior, marking the onset of the development of a biofilm: loss of flagella or cilia, changes in cell phenotype and start of reproduction process to create a microcolony.

\begin{figure}
\centering
\includegraphics[width=10cm]{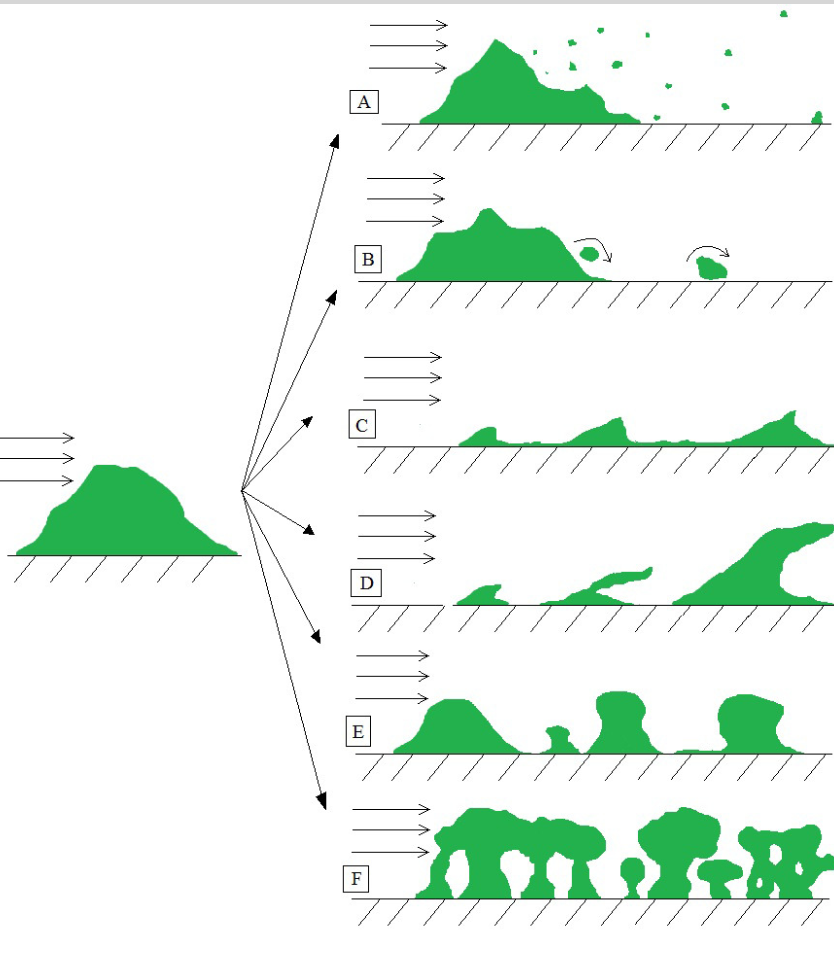}
\caption{ Expansion and pattern formation mechanisms in submerged biofilms.
(a) Darting: bacteria are expelled from the colony.
(b) Rolling: bacterial clusters detach and roll over the surface. 
(c) Rippling: ridges and valleys form on the biofilm surface and may travel downstream. 
(d) Streamers: fingers curved in the direction of the flow.
(d) Mounds: groups of hills. 
(e) Mushroom networks: mushroom-like towers grow and merge creating a porous 
structure with channels.}
\label{fig0}
\end{figure}

If aerobic bacteria receive sufficient amounts of carbon and oxygen, they divide and proliferate on the surface forming microcolonies that constitute the germ of a biofilm and may eventually merge.  The metabolism of the cells near the biofilm surface restricts the diffusion of oxygen and carbon to bacteria at the bottom of the colony. A fraction of cells in the biofilm begin to produce  exopolysaccharides, forming the EPS matrix. This promotes vertical growth of the colony, improves access to oxygen and carbon, and gives rise to a macrocolony \cite{development,cooperation}. Cells therein are held together by a matrix that also contains dead cell debris and extracellular DNA. The surrounding flow subjects cells in the upper part of the colony to higher shear forces and these cells may be detached from the colony.  External conditions modify its main characteristics, generating different internal and external structures optimized to improve nutrient obtention and mechanical resistance to external shear stress. Typical macrocolonies consist of mushroom-like towers separated by fluid-filled voids carrying nutrients and oxygen, although flat structures are also possible.  Depending on the hydrodynamic conditions, the nutrient availability, the nutrient source and the bacterial strain, circular colonies, streamers, ripples, rolls, streamlined patches or mushroom networks  \cite{paureaginosa, stoodleyfigures} are observed, see Figure \ref{fig0}. A reversible evolution from one pattern regime to another by increasing or decreasing the nutrient concentration and the shear force has been observed experimentally in \cite{stoodleycohesion}.

Once the biofilm has reached a certain thickness and its structure is well defined, quorum sensing processes activate the spread mechanism. This mechanism consists of a set of different strategies to release bacteria to the flow in order to colonize new surfaces and ensure the strain survival \cite{bryers,stoodleyfigures}. Some examples are structure self weakening in the form of flow detachment or surface expansion produced by dragging mechanisms, like rippling, rolling or darting \cite{development, paureaginosa}. { Figure
\ref{fig0} illustrates some of these mechanisms.}

If the nutrient or oxygen concentration is not enough to sustain the population, cells start dying until an equilibrium between deaths, availability of nutrients, cellular growth and spread  is reached.

{ Series of experiments (see \cite{paureaginosa} and \cite{stoodleyfigures, stoodleynutrients,stoodleyripples,stoodleysulfate,stoodleycohesion}, for instance) were performed to assess the influence of hydrodynamic conditions and nutrients on biofilm structure for specific bacterial strains.  Experiments were made in rectangular glass flow cells with sections of a few millimeters. We summarize a few conclusions that will be kept in mind in the sequel. The EPS matrix was shown to determine biofilm cohesive strength.  The shear force created as the fluid flows over its surface seemed to be the principal physical force acting on the biofilm. Biofilms grown under higher shear were observed to adhere more strongly and have a stronger EPS matrix than those created at low shear. Increasing the velocity of the fluid, biofilms created at lower Reynolds numbers were washed out. At low shear, detachment takes the form of sloughing. For higher shear forces, detachment occurs in the form of erosion, producing smoother, flatter and thinner biofilms.
Mounds, ripples and streamers were identified. As the Reynolds number grew, round biofilm patches  elongated with the current. The drag seemed to push newborn cells downstream.  While ripples migrated downstream sliding over lower layers of biofilm without detaching, streamers seemed to be anchored.} 
{ Ripples were observed in laminar and turbulent flows. Streamers were believed
to appear in turbulent regimes.
Recent experiments carried out in curved pipes with sections of a few hundred microns \cite{stone} have produced streamers in laminar flows, raising new
issues about the mechanisms leading to streamer and pattern formation in submerged biofilms.}

\begin{figure}
\centering
\includegraphics[width=8cm]{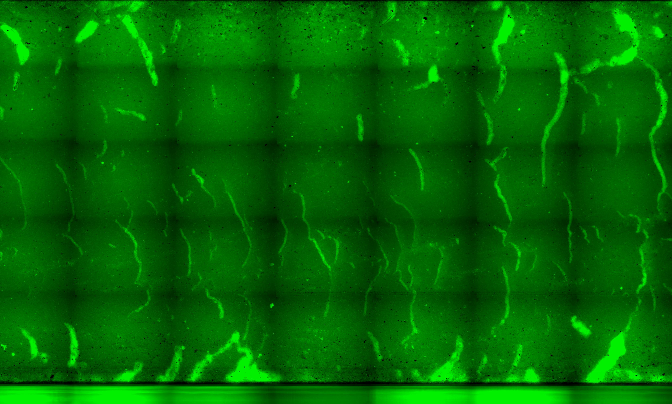}
\caption{Panoramic view of ripples formed by Pseudomonas putida on the bottom of a duct at $Re=1000$, obtained by glueing together images taken along a grid.}
\label{fig5}
\end{figure}

{ Additional experiments in rectangular flow cells, but on polycarbonate surfaces,  consider the influence of roughness \cite{unpublished}. Comparing biofilm grown on standard surfaces and milled surfaces with an average roughness about  $2-4$ $\mu$m (slightly larger than the size of bacteria), more biofilm was observed to accumulate on the milled surface. Increasing the Reynolds number, almost no biofilm accumulated on the untreated surface. Patches, uniform covers with mounds, and ripples were identified on the milled surface as the Reynolds number was increased, see Figures \ref{fig4} and \ref{fig5}. The experiments suggested that the specific roughness pattern selected might enhance biofilm formation and expansion.}

\section{Two dimensional model }
\label{overall}

\begin{figure}[!t]
\centering
\includegraphics[width=10cm]{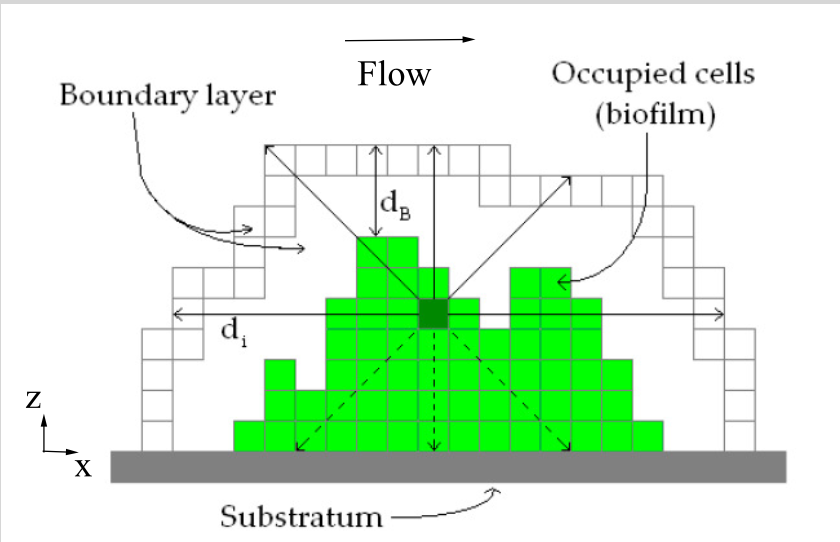}
\caption{Geometry for the cellular automata description.}
\label{fig2}
\end{figure}

In line with the experiments described at the end of the previous section,
the geometry selected for our model is a rectangular duct with a rough bottom 
surface. The roughness has the same order of magnitude as the bacterial size. 
An aqueous solution containing nutrients flows along the pipe in the $x$ direction. 
We take a longitudinal section of the rectangular pipe ($x$=length, $z$=height) and 
study biofilm evolution in it, see Figure \ref{fig2}. {Next section extends the model
to three dimensions.}

We consider bacteria as living entities that may perform a certain
number of activities depending on external factors.  They may divide and spread,
generate EPS matrix,  deactivate, detach from the biofilm  and so on.  Aerobic bacteria 
need oxygen and a carbon source to survive. Bacteria choose to
carry out one activity or another with a certain probability according to the levels of 
nutrient and oxygen available at its location and the shear force exerted by the flow. 
This point of view makes it easier to incorporate the interaction with roughness
patterns. { In most simulations, biofilms will be grown from an initial seed that is already 
attached to the substratum. We will allow for adhesion of floating cells at the end.}

\begin{figure}[!t]
\centering
\includegraphics[width=8cm]{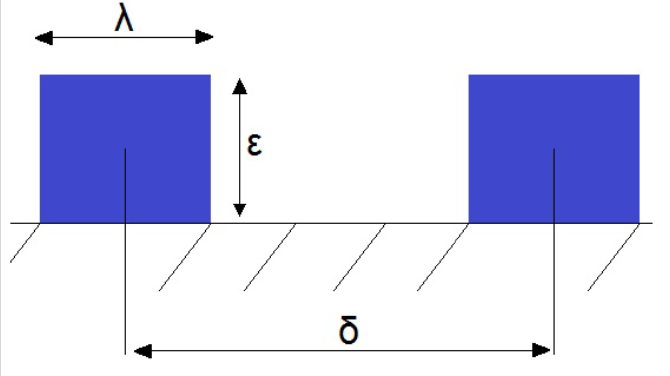}
\caption{ Roughness pattern.}
\label{fig24}
\end{figure}

In a CA model, space is represented by a grid of square tiles, see \cite{hermanovic,qca} and references therein.   In view of our roughness patterns,  we choose the size of the tiles $a$ to be similar to the size of one bacterium (microns) \cite{nota}.  Notice that we envisage tubes with sizes of a few hundred microns.
Each tile is filled with either one bacterium, water or bottom surface material, and its status may change from one time step to the next according to the rules governing the different processes. Each individual bacteria evolves according to the nutrient and oxygen presence it feels, to its affinity to the carbon source,  to the flow strength it feels and to its location in the biofilm.
Roughness is modeled by rectangular steps on the bottom, characterized by their height $\epsilon$, the length of the peaks $\lambda$, and the distance between them $\delta$, see Figure \ref{fig24}.

We  propose a modular CA model  organized in sub-models that depend on the aspects of biofilm formation and evolution we want to describe. Each sub-model  deals with a distinctive bacterial mechanism: cellular division and spreading to neighboring tiles, generation of EPS matrix { and decay, cell detachment
and adhesion}. 
For each individual bacteria each of these events  will be assigned a probability that usually depends on the concentration of oxygen and nutrients, the fluid flow, the biofilm cohesion or the number of neighboring cells containing biomass and their location. We have chosen to cover the following aspects of biofilm formation and evolution:
\begin{itemize}
\item Dynamics of dissolved components (nutrients, oxygen) outside the biofilm  and inside.  Experiments are usually designed in such a way that their concentrations are almost constant outside the biofilm. Inside, they are governed  by reaction-diffusion equations, but a quasi-steady approximation thereof might suffice because the diffusion and reaction of dissolved components are likely to be faster than the rates of biological processes \cite{hybridmatrix,steady,3d}.  
\item Detachment of isolated bacteria, biofilm erosion and detachment of biofilm fragments. Each cell has a probability to detach from the biofilm depending on its location, the number and location of neighboring bacteria,   the force exerted by the flow  and the cohesion of the biofilm. Detached bacteria are carried by the flow. Biofilm fragments attached to the rest of the film by a few cells  may also be eroded if the connection to the substratum breaks off.
\item  Generation of EPS matrix { and decay}. Each bacterium has a probability to produce EPS matrix depending on the flow and the concentration of substrate and oxygen \cite{matrix1,matrix2,cooperation}.  The cohesion of the biofilm depends on the EPS matrix, which in turn affects mechanisms such as detachment
\cite{stoodleycohesion}. { Cell deactivation and decay mechanisms for low concentrations may generate inerts playing a role in biofilm adhesive properties.}
\item Reproduction and spreading. Each bacterium has a probability to reproduce depending on the availability of oxygen and nutrients. New bacteria fill neighboring empty tiles or shift existing bacteria with a certain probability. 
\item { Adhesion of floating cells. The bacteria carried with the flow may attach either to existing biofilm parts or to uncolonized parts of the substratum with a certain probability   that depends on the flow and the affinity between the bacterial strain and the surface \cite{adhesion1,adhesion2}.}
\end{itemize}
{ Selecting these basic mechanisms we are able to generate patterns and reproduce behaviors that remind of those observed in real biofilms. The true biological
processes being yet largely unknown, the idea is to choose simple rules motivated
by experimental observations trying to mimic some observed behaviors.
Whenever a better understanding of the cellular processes is available, the proposed rules can be updated to reflect that knowledge.}
{ The   selection of processes considered here provides  insight on the role of different parameters in the structure of cellular aggregates, together with basic understanding of the way some competing cellular mechanisms may act and interact. This information might be incorporated in more refined models, or used for calibration to experiments.
}
In our present model, the flow influences detachment { and attachment} processes, the concentration boundary layer outside the biofilm, and EPS matrix generation and cell reproduction through it. In our simulations, the biofilm/fluid interface moves, but as a result of erosion, adhesion, mass  production and spreading. Depending on the shear force and the biofilm cohesion, the fluid might also move  biofilm blocks or cells. { A mechanism allowing external forces to shift cells is yet to be included.}  

We describe below our approach to incorporate the above mentioned mechanisms in each time step
of the biofilm evolution. 

\subsection{Dissolved components}
\label{dissolved}

The evolution of an  aerobic  biofilm depends on the availability of carbon sources and oxygen. The concentrations of oxygen $c_o$ and substrate $c_s$ are governed by a convection-reaction-diffusion system. Since the diffusion and reaction of dissolved components are faster than the rates of biological processes \cite{steady,hybridmatrix,3d},  quasi-steady approximations suffice. Outside the biofilm, the concentrations follow uncoupled convection-diffusion equations. Inside the biofilm, the convection due to the flow disappears, but coupling reaction terms representing nutrient and oxygen uptake by cells must be included.  The model is completed with boundary conditions at the walls of the duct and the bulk/boundary layer interface \cite{3d}.  
Experiments are usually designed to keep concentrations almost constant within the bulk fluid.  Nutrient concentration gradients result from a combination of nutrient transport from the bulk fluid through a concentration boundary layer adjacent to the biofilm/fluid interface and nutrient uptake uptake by the cells. The thickness of the concentration boundary layer characterizes the external mass transport and depends on the flow regime. Experimental measurements of the nutrient concentration gradients and the boundary
layer thickness for different flows are performed in \cite{blayer,blayer2}. The thickness $d_B$ was shown to depend on the bulk velocity of the flow in \cite{blayer} (it seems to be inversely proportional to it). Additional experimental studies show that transport outside cell aggregates is larger than inside them \cite{blayer3}. Outside the boundary layer, advection dominates the transport, while inside the aggregates diffusion is the controlling factor. This motivates the assumption  that transport through the cellular aggregate and the boundary layer occurs by diffusion with the same effective diffusion coefficient for the boundary layer and the biofilm \cite{blayer3,hermanovic}.

Choosing the oxygen $C_{o}$ and nutrient $C_{s}$ concentration at the bulk/boundary layer interface and the  boundary layer thickness $d_B(Re)$ as control parameters, the concentrations of nutrients   and oxygen   inside the region containing the biofilm and the boundary layer \cite{hermanovic} are governed by:
\begin{eqnarray} 
D_s \Delta  c_s =k_2 {c_s\over c_s+K_s} {c_o\over c_o+K_o}, \label{cs}\\
D_o \Delta  c_o =\omega k_2 {c_s\over c_s+K_s} {c_o\over c_o+K_o},  \label{co}
\end{eqnarray}
with zero flux conditions at the substratum. 
The right hand sides represent the nutrient and oxygen uptake kinetics. Here, $\omega$ is the stoichiometric coefficient of the oxygen reaction, $k_2$ the uptake rate of the nutrient, 
$D_o$ the diffusion coefficient for oxygen, $D_s$ the diffusion coefficient for the substrate,
$K_o$ the Monod  half-saturation  coefficient of oxygen, and $K_s$ the Monod  half-saturation  coefficient of the carbon source. The values of all these parameters depend on the bacteria species forming the biofilm. The use of Monod laws makes sense provided the values of the concentrations remain small. Otherwise, inhibition terms should be incorporated. { We assume that oxygen is in excess, which is often the case. The concentration of nutrients $c_s$ becomes the limiting concentration $c_l$, that is, the one that penetrates a shorter distance  into the biofilm and therefore constrains division and survival of cells and biofilm growth thereof. The system is reduced to: 
\begin{eqnarray}
D_s \Delta  c_s =k_2 {c_s\over c_s+K_s}. \label{csred}
\end{eqnarray}
}
For submerged biofilms  nutrients and oxygen are both provided by the  surrounding flow. Biofilms grown on air/solid surfaces \cite{kolter} might take oxygen from the air and the substratum, and nutrients from the substratum. In those cases, oxygen and nutrients might become limiting factors in different regions.  

At each time step, (\ref{csred}) is solved to reflect changes of nutrient uptake caused by the new biofilm geometry. In our tests, we have computed numerically the solution of the nonlinear boundary value problem for the concentration  by using an iterative relaxation scheme with local error control.  Solutions for the elliptic problem are constructed as stationary solutions of the diffusion problem. In the tests presented here,  error tolerance is set to $10^{-3}$. A simple analytical formula creating a concentration distribution inside the biofilm, adapted to its boundary and with a reasonable qualitative dependence on the parameters, was proposed in \cite{hermanovic}. That formula may produce useful qualitative predictions at a much lower computational cost if plugged into our stochastic description.


\subsection{Erosion by the flow}
\label{detachment}

Surface cells are subject to shear forces exerted by the flow, which may detach them from the biofilm \cite{stoodleyfigures,stoodleynutrients}. In principle, cells sheltered by other cells are somehow protected from erosion.  Exposed cells will detach with a probability depending on the number and location of their neighbors relative to the motion of the fluid, the biofilm cohesion, which is controlled by EPS matrix generation,  and the force due to the flow felt by them, which depends on the Reynolds  number.   The Reynolds number $Re$ is computed using the hydraulic diameter of the ducts and the average velocity, which are known. 

\begin{figure}[!t]
\centering
\includegraphics[width=5cm]{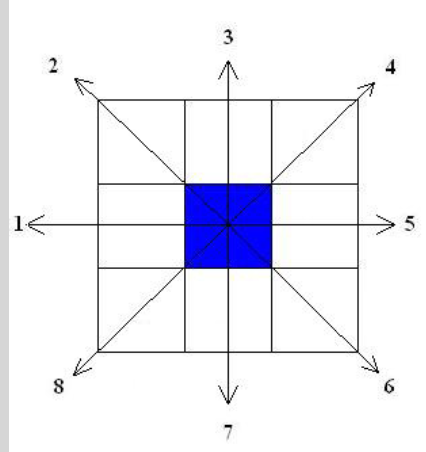}
\caption{Neighbor location in the cellular automata description.}
\label{fig3}
\end{figure}

The total force acting on a cell is a balance between the
force exerted by the flow and additional forces exerted by its neighbors, depending on their distribution and on the biofilm cohesion. According to the geometry depicted in Figure \ref{fig2}, the force due to the flow is mostly oriented in the $x$ direction.  Figure \ref{fig3} illustrates the location of possible neighbors. Each of them adds a force with a component opposite to the direction of the flow. The strength of each force depends on the position of each neighbor. 
Let us number the eight neighboring tiles clockwise, starting with the western direction, so that $n_1,n_2,n_3,n_4,n_5,n_6,n_7,n_8$ denote the neighbors located to the west, northwest, north, northeast, east, southeast, south, and southwest, respectively.  If the cell has a western neighbor $n_1$, it is partially shielded from the flow and is unlikely to be carried away. The strongest resistance against the flow is exerted by the eastern neighbor $n_5$. Next  in magnitude are the resistance forces due to adjacent cells $n_4$ and $n_6$ located in the northeast and southeast directions, and then $n_3$ and $n_7$ in the north and south directions. Neighbors $n_2$ and $n_8$ in the northwest and southwest directions add little resistance.

{The magnitude of the local force acting on a cell ${\cal C}$ can be described by:
\begin{eqnarray}
\tau({\cal C})= \tau(Re) (1- \beta(Re) \chi_1({\cal C})) \left(1 - \sum_{i=2}^8 e_i \chi_i({\cal C})\right). 
\label{shearmod}
\end{eqnarray}
 Here, $\tau(Re)$ represents the shear force due to the flow. Solving Navier-Stokes equations outside the biofilm at each step to evaluate it is too costly, especially in turbulent regimes. We  are studying small biofilms growing at the wall. Therefore,
we approximate it by the shear force at the bottom of an unperturbed rectangular tube. 
The shear stress at the substratum surface is evaluated from the velocity profiles as 
$\tau=-\mu {du \over dz}$, where $\mu$ is the fluid viscosity. For laminar flows, $\tau(Re)$ is 
known explicitly: $u$ is given by the Hagen-Poiseuille expression. 
In general, it can be estimated as 
\[
\tau(Re)= {f \rho  u^2 \over 2},
\]
where $u$ is the known average flow velocity, $\rho$ the fluid density
and $f$ the friction factor given by ${16 \over Re}$ for laminar flows
and ${0.0791\over Re^{0.25}}$ for turbulent flows (see \cite{shear1,shear2}).
Notice that the roughness Reynolds number  $Re_r={\rho  u  e  f^{0.5} \over \mu
8^{0.5}} $ is expected to be small, $e$ being the height of the roughness elements.



$\tau(Re)$ is multiplied by a factor that takes into account the geometry of the biofilm
and the local support provided by neigbouring cells depending on their distribution.
The functions $\chi_i({\cal C})$ in (\ref{shearmod}) take the value $1$ whenever the cell ${\cal C}$
has a neighbor located at the position $n_i$, and vanish otherwise. The factors
and weights have been chosen to account for the  fact that the fluid flows in the $x$ direction. 
The weights satisfy $e_i\in (0,1)$, $\sum_{i=2}^8 e_i=1$, {and $\beta(Re) \in (0,1)$, so that 
the sign of $\tau(Re)$ is not reversed.}
In our computer experiments, we have set $e_5={5 \over 17}$, $e_4=e_6={3 \over 17}$, 
$e_3=e_7={2\over 17}$, $e_2=e_8={1\over 17}$. They represent the
added resistance against the flow due to neighboring cells depending on their position.
{$\beta(Re)$ takes values close to $1$. If we set $\beta(Re)=1$, $\tau({\cal C})=0$
whenever the cell has a western neighbor. As it departs from $1$, the probability
of a cell being eroded in presence of the western neighbor $n_1$ grows.
This may be more likely as $Re$ increases.  }

A probability for cell erosion can be defined following \cite{hermanovic}:
\begin{eqnarray}
P_e({\cal C}) = {1 \over 1 + { \sigma({\cal C}) \over   \tau({\cal C})}}
=  {\tau({\cal C}) \over \tau({\cal C}) +  \sigma({\cal C})}.     \label{erosion}
\end{eqnarray}
Whenever $\tau({\cal C})=0$, we set $P_e({\cal C})=0$. Here, $\tau({\cal C})$ is given by
(\ref{shearmod}), and $\sigma({\cal C})$ represents the biofilm strength. 
A cohesion parameter $\sigma({\cal C})$ varying in accordance with the local EPS
generation production is introduced in Section \ref{matrix}.  
 At each time step and for each cell, we generate a random number $r \in(0,1)$.
When $r<P_e({\cal C})$, the cell detaches from the biofilm. Erosion due to the flow may occur  
as detachment of single cells or of whole clusters of bacteria with a thinning connection to the 
rest of the biofilm. 
{ In Figures \ref{fig81}-\ref{beta}, $\sigma({\cal C})$ is set equal to a constant, to investigate the influence on the biofilm structure of the erosion mechanism we propose. In Figure  
\ref{fig810}, $\sigma({\cal C})$ varies as described in the next section.} 


While programming the erosion mechanism, we chose to let two cells 
be connected if they are  neighbors in the grid via any of the $8$ 
neigbouring directions. This is  done for several reasons. 
First, for consistency with division, because the cells divide in each of 
the 8 directions. Further, if we would only let the 4 main directions
connect cells, then newly formed cells with only diagonal
connections would break off immediately. 
We have coded both versions, and simulations show that qualitatively
the behavior of the system is unchanged. The difference is that
with connections in only 4 main directions the erosion is a little
stronger for reasons mentioned above. 

\subsection{EPS matrix generation}
\label{matrix}

In small colonies, most bacteria reproduce again and again. As the size of the biofilm grows, the density of cells increases and nutrients and oxygen become scarce. 
Bacteria start to produce EPS matrix with a certain probability \cite{development}. 
EPS generation is believed to push newborn cells upwards, inducing vertical growth and making easier access to food and oxygen for them \cite{development}, 
see Figure \ref{fig1}. The EPS matrix also spreads over the neighbouring bacteria making their reproduction harder. As bacteria are deeper in the biofilm, their chances to produce EPS seem to increase because of low concentrations.

{ The nature of the surrounding flow also influences EPS generation. The stronger the shear due to the flow is, the more resistant the EPS matrix is \cite{stoodleyfigures,stoodleycohesion}. Biofilms grown at fixed Reynolds numbers tend to be carried away with the flow as the Reynolds number is increased.}

{ According to these observations, we assume that the probability of a cell producing EPS matrix  depends on the availability of nutrients and oxygen at the cell position and the shear exerted by the flow.  The EPS matrix is generated with the following probability law:
\begin{eqnarray}
P_{eps}({\cal C}) = R(Re) \left(1-{c_l({\cal C}) \over c_l({\cal C}) + K_l}\right), \label{eps}
\end{eqnarray}
where  $R(Re) \in (0,1)$  and $c_l$ represents the limiting concentration.   Cell decay may be taken care of by deactivating cells at which the concentration falls below a critical value.}
Concentrations at each cell ${\cal C}$ are stored in a linked list of cells. These lists are updated once a time step has been completed, to reflect the change in  the biofilm geometry. 
The concentrations are computed as described in Section \ref{dissolved}.  

{ As expected \cite{development}, the probability (\ref{eps}) takes small values when the biofilm thickness does not surpass a threshold (the limiting concentration is large enough) and increases as  the cell is deeper into the biofilm (the limiting concentration will decrease). }

{ At each time step and for each cell, we generate a random number $s \in(0,1)$. When $s<P_{eps}({\cal C})$, the cell will generate EPS. After several steps, the fraction of cells generating EPS matrix stabilizes to a certain value, that may be used to determine $R(Re)$ by comparison with experimental measurements. Whereas $R(Re)$ controls the percentage of cell generating EPS matrix, the factor involving the concentration governs the spatial distribution of these cells in the biofilm.}

{ The amount and nature of the EPS matrix produced determines the cohesion (strength) of the biofilm.  
Parameters representing the biofilm cohesion can be measured \cite{stoodleycohesion,
stone} and introduced in the erosion probability law (\ref{erosion}). However, it may
be useful to have a rough idea of its spatial variations and their effect, to infer how the presence of weaker regions may affect biofilm evolution or to input this information
in macroscopic models.}
 
{The  EPS matrix diffuses and accumulates in different ways in different biofilms \cite{matrix1,matrix2}. 
We test here a  local measure $\sigma$ of the biofilm cohesion which takes into account the number of neighbors and their nature:}
\begin{eqnarray}
\sigma({\cal C})= {\sigma_0(Re) \over 8} \sum_{i=1}^8  \sigma_i({\cal C}), 
\label{sigma} 
\end{eqnarray}
where
\begin{eqnarray}
\sigma_i({\cal C})= \left\{ \begin{array}{ll}
0 & \mbox{if neighbor}  \; n_i \;  \mbox{is not present }\\
\alpha & \mbox{if neighbor} \; n_i \; \mbox{is present, but does}\\
 & \mbox{ not produce EPS matrix}\\
1 & \mbox{if neighbor}  \; n_i \;  \mbox{produces EPS matrix} 
\end{array} \right.
\end{eqnarray}
and $n_1({\cal C}),$  $n_2({\cal C})$,...,$n_8({\cal C})$ denote the eight neighbor locations for the cell under study (see Figure \ref{fig3})  and $\sigma_0, \alpha >0$. { When 
deactivated cells are present an additional constant $\alpha'$ should be
used for them.}
{Other options are possible.  $\sigma({\cal C})$ might be modulated 
by the variations of the concentration of EPS matrix, governed by equations
similar to those described in Section \ref{dissolved}   when the matrix 
diffuses easily.}

The parameters $\sigma_0, \alpha$ represent the strength of the EPS matrix generated 
by the bacteria and the strength of the attachment between standard 
bacteria.  In practice, $\sigma_0$ seems to increase with $Re$  \cite{stoodleyfigures,stoodleycohesion,detachment}.  
They  depend on the type of bacteria and must be fitted experimentally.   
We have selected   $\alpha={1 \over 2}$ in our computer experiments. 
Figure \ref{fig810} includes the variable cohesion in the erosion mechanism. 

{ For specific bacteria forming biofilms on air/agar interfaces in absence of flow,
there are detailed measurements of the fractions of cells generating EPS and related
chemicals. Precise visualizations of their spatial distribution within the biofilm are 
available too \cite{kolter}. Regions with large availability of nutrients contain normal cells.  
As the concentration of nutrients decreases, the percentage of cells generating EPS 
matrix increases, and they may even deactivate.  In that specific case, EPS matrix
production is known to be triggered by cell production of several chemicals.
We are not aware of such detailed studies for biofilms in flows yet, but the
laws described here might be updated to incorporate such knowledge
if it ever becomes available.}

\subsection{Cell division}
\label{reproduction}

The mechanism for cell reproduction is similar to that in \cite{hermanovic}, except  
for the fact that EPS producers { and deactivated cells } do not undergo cellular division 
{ in the same step.}
At each time step, and once we have checked which cells produce EPS 
matrix, the remaining cells ${\cal C}$ will divide  with probability:
\begin{eqnarray}
P_d({\cal C}) ={c_l({\cal C}) \over c_l({\cal C}) + K_l},
\label{division}
\end{eqnarray}
where $c_l$ denotes the limiting concentration and $K_l$ its saturation coefficient in the 
Monod law. The concentration is computed at the beginning of each step as described
in Section \ref{dissolved}.  We are neglecting changes in concentration due 
to newborn cell consumption or cell switching to EPS generation within the same step.

At each time step, and for each cell not generating EPS matrix,
we compute a random number $p \in (0,1)$. If $p<P_d({\cal C})$,
the cell will divide.  Whenever neighboring grid tiles are empty, the daughter
cell is placed in any of the empty tiles with equal probability.
Otherwise, the new cell will shift one of the neighbors. The cell
offering the minimal mechanical resistance is chosen, that
is, the one lying in the direction of shortest distance from 
the reproducing cell to the biofilm boundary/bulk layer.   
The same rule should be applied to the shifted cell: it either occupies adjacent 
empty tiles with equal probability or shifts a neighboring cell in the direction
of smallest mechanical resistance. However, to reduce the computational
cost we shift neighboring cells in the same direction. This process is 
repeated until all the displaced cells have been accommodated.

\subsection{Adhesion of floating cells}
\label{sec:adhesion}

{ The adhesion mechanism assumes that we know the number of bacterial cells
that are floating in the flow and the adhesion rate. 
Let $N_f(t)$ be the number of cells carried by the flow at time step $t$. The number 
of cells that will attach to the bottom surface at time $t$ will be a fraction of the number
of floating cells, larger or smaller depending on the Reynolds number, the type of 
surface and the bacteria species. Raising the Reynolds number we increase the
probability of hitting the surface. Whether the bacteria successfully attach will
rely on the interaction between the specific type of bacteria and surface we are
working with. Thus, the number of attached cells will be:
\begin{eqnarray}
N(t)= [\gamma N_f(t)], \label{attached}
\end{eqnarray}
where $[ \,]$ denotes the integer part and $\gamma$ is a parameter  which measures the likeliness of that specific bacteria to attach to that surface. It can be seen as an adhesion rate.
Bacterial likeliness to attach to a surface depends on the type of flow and the nature of the substratum \cite{adhesion1}. In laminar flows  the main mechanism 
driving particles to the wall seems to be Brownian motion \cite{pedro}. This usually results in low deposition rates for laminar flows. As the Reynolds number increases, turbulent 
effects play a role and particle deposition rate increases linearly with the Reynolds number for small Stokes numbers  \cite{pedro}.  It has been experimentally observed 
that the residence time of bacteria hitting a wall increases with shear
\cite{adhesion2} and that biofilm accumulation tends to be larger for larger flows \cite{stoodleysulfate}. Thus, $\gamma(Re)$ is likely to increase with $Re$.

To decide where these cells are going to be attached, we assign a number to any surface compartment (of either substrate or biofilm, both are considered together), producing a list of $S$ numbers. Then, we generate $N$ random integers between $1$ and $S$. We locate new cells at those  positions. This assumes equal probability for all the surfaces as adhesion sites. Numerical tests implementing this
mechanism are shown in Figures \ref{adhesion1} and \ref{adhesion4}.
Whenever precise information on preferential adhesion sites is available, as in \cite{stone}, the adhesion strategy should be changed to account for that fact.}

\subsection{Nondimensionalization and parameters}
\label{parameters}

Nondimensionalizing the model is essential to identify the minimum number of independent parameters, or to be able to distinguish what is large and what is small. All the probabilities we have introduced are dimensionless. Dimensions enter the model mainly through the concentration, length and time scales:

\begin{itemize}
\item Length: The basic distance considered in the model is the size of a bacteria $a$, about $2$ or $1$ micrometers. It is set equal to $1$ in our tests.

\item Time: In the model, time is not given explicitly. It appears in the number of time steps carried out at each simulation. A simple estimate for the time step size can be given: in the most favorable conditions for bacterial reproduction, the concentration is so high that the probability of reproduction is approximately one. In these conditions, the bacterial population will double in a single time step. An upper bound for the time step,  which allows to relate computational $T$ and experimental times is the minimum doubling time:
$$ t=\frac{\ln(2)}{\nu_{max}}, $$ 
where $\nu_{max}$ is   the growth rate, which is a known parameter for 
some bacterial species and nutrients.

\item Concentration: The concentration field is calculated solving a boundary value problem for the limiting concentration, which involves a number of constants with their units that must be nondimensionalized. Making the changes of variables:
\[ \hat{c}_l\!=\!\frac{c_l}{K_l}, \, \hat{C_l}\!=\!\frac{C_l}{K_l}, \; F_l \!=\!\frac{k_l a^2}{2D_l K_l}, \; \delta_B\!=\!\frac{d_B}{a},  \hat{x}\!=\!\frac{x}{a},
\]
we get the dimensionless equations:  
\begin{eqnarray}
\hat{\Delta}  \hat{c}_l = 2F_l {\hat{c}_l \over \hat{c}_l + 1}. \label{dimensionless}
\end{eqnarray}
with boundary condition  $\hat{C}_l$ at the bulk/boundary layer interface.
\end{itemize}

The four parameters $a$, $k_l$, $D_l$, $K_l$ are reduced to one: $F_l$,  
{ which is analogous to the Thiele modulus and measures the ratio
of the uptake rate to the diffusional supply.} Typical values for bacteria commonly 
used in flows, like Pseudomonas Aureaginosa
or Pseudomonas Putida, and standard nutrients produce values in the range
$10^{-8}-10^{-2}$.
Notice that in all the formulas introduced in previous sections, only quotients 
${c_l \over c_l + K_l} = {\hat{c}_l \over \hat{c}_l + 1}$ are involved.  All combinations 
of parameters producing the same value $F_l$ for fixed values of the remaining
parameters produce the same results.


The tables below summarize the different parameters used through the
text. For fixed bacterial strains, nutrients, surfaces and flows, the 
parameters $a$, $\varepsilon$, $\lambda$, $\delta$,  $C_s$, $C_o$, 
$D_s$, $D_o$, $\omega$,    $Re$, $\rho$, $\mu$, $N_f(t)$ are usually known. 
The parameters describing the bacterial kinetics for the particular
choice of nutrient  $k_2$, $K_s$, $K_o$, $\nu_{max}$ are only
available in some cases. In general, they have to be measured.
{ The same happens with the average cohesion $\sigma_0(Re)$
or the adhesion rate $\gamma(Re)$.}
The boundary layer thickness $d_B(Re)$ may be estimated
experimentally for a given flow regime. $\sigma_0$ and $\alpha$
in (\ref{sigma}) might be calibrated using experimental measurements on 
the percentage of biofilm cells generating EPS matrix and its cohesion.
The  specific values of $e_2, ..., e_8$ in (\ref{sigma})
are not too relevant. { Replacing them by other
positive values   respecting the symmetry
produces similar results}.  
The dependence on temperature in the model
is implicit through the uptake rates, the density and viscosity of the fluid,
and the diffusivities.

\section{Three dimensional model}
\label{3dmodel}

\begin{figure}[!ht]
\begin{center}
\includegraphics[width=6cm]{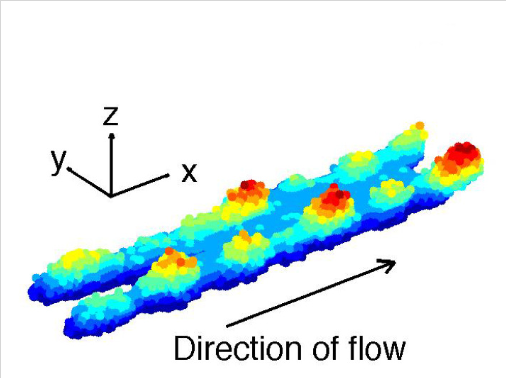}
\end{center}
\caption{Three dimensional biofilm growing on the bottom of the pipe.}
\label{fig30}
\end{figure}

\begin{figure}[!ht]
\begin{center}
\includegraphics[width=10cm]{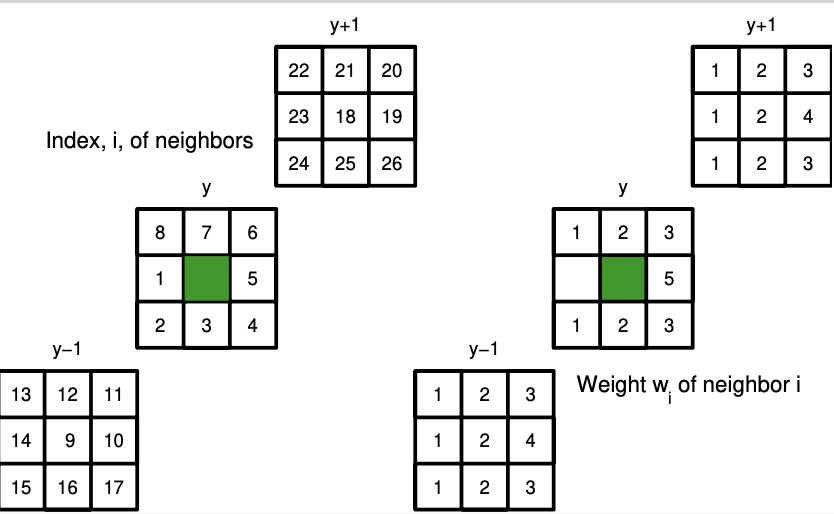}
\end{center}
\caption{Neighbors and weights for three dimensional biofilms.}
\label{fig31}
\end{figure}

{ The model can be extended to three dimensions with simple
changes. Biofilms grow on the bottom of a rectangular
pipe carrying a flow containing nutrients, as shown in Figure \ref{fig30}.
Space is partitioned in a grid a cubic tiles, of size $a$. Again,
each of them is filled with one bacterium, fluid or substratum.
Bacteria reproduce, spread, detach, decay and generate EPS matrix
according to the rules described in Section \ref{overall}, with the
changes we describe below.

The probabilities for EPS generation and cell division are
still given by (\ref{eps}) and (\ref{division}), respectively.
Quotients ${c_l \over c_l + K_l} = {\hat{c}_l \over \hat{c}_l + 1}$
are involved, where the dimensionless limiting concentration field
is computed by solving numerically the three dimensional version
of the nonlinear boundary value problem for the concentration. We use
a relaxation method to compute numerically the solutions.

The erosion probability has the form (\ref{erosion}), where $\tau({\cal C})$
is given by (\ref{shearmod}) with  $\sum_{i=2}^8 e_i \chi_i({\cal C})$
replaced by $f({\cal C})$:
\begin{eqnarray}
f({\cal C}) = {1\over 2} f_y({\cal C}) + {1\over 4} (f_{y+1}({\cal C})+f_{y-1}({\cal C})), 
\hskip 5mm \label{shearmod3D} \\
f_{y-1}={1\over 19} \sum_{i=9}^{17} w_i \chi_{i}({\cal C}), \quad
f_{y+1}={1\over 19} \sum_{i=18}^{26} w_i \chi_{i}({\cal C}),  \quad
f_{y}= {1\over 17} \sum_{i=2}^8 w_{i} \chi_{i}({\cal C}). \nonumber
\end{eqnarray}
$\tau(Re)$ represents the shear at the bottom wall. This
assumes that the patterns remain near the wall. Explicit formulas for laminar
flows are available in ducts with rectangular section \cite{laminar3d}, and approximations 
for turbulent flows can be found in cite \cite{laminar2d}.
Numerical simulations in Figures \ref{washout}-\ref{mound} set $\tau(Re)$
equal to a constant to simplify. 
The functions $\chi_{i}({\cal C})$ take the value one whenever the cell has a neighbor located 
at the position $n_{i}$, respectively, and vanish otherwise. $n_{i}$, $ i=1,..,26$, denote
the neighbors of the cell under study $n_0$, see Figure \ref{fig31}.  The weights  $w_{i}$ 
are chosen to account for the fact that the fluid flows mostly in the $x$ direction. 
In our simulations, the weights $w_i$ are slightly larger for neighbors located in the 
intermediate slice of the cube, which contains the cell whose neighbors we are tracking.
The weights are equal for neighbors occupying the same position in each of the two 
lateral slices. Figure \ref{fig31} illustrates the numbering of neighbors and the 
weights we have used in our simulations.  
The local cohesion in (\ref{erosion}) may be constant or take the form:
\begin{eqnarray}
\sigma({\cal C})={1\over 3}(\sigma_y({\cal C})+\sigma_{y-1}({\cal C})+\sigma_{y+1}({\cal C})),          
\label{sigma3D}
\end{eqnarray}
where $\sigma_y({\cal C})$ is given by (\ref{sigma}) applied to the slice
containing the cell. $\sigma_{y+1}({\cal C})$, and $\sigma_{y-1}({\cal C})$
are given by a similar formula, but replacing $8$ by $9$ and summing
up over all the neighbors in the lateral slices.}

\section{Numerical results}
\label{numerical}

{In this section, we illustrate the evolution of several initial configurations under 
different conditions, to gain understanding of the influence of the controlling
parameters and the interaction between competing mechanisms. }
First, we analyze the evolution of a biofilm seed for constant biofilm cohesion
$\sigma$, considering only the growth and erosion mechanisms for different shear 
and nutrients on either flat or rough surfaces. {Then, we implement the adhesion 
mechanism on uncolonized surfaces.}
Finally, we include the EPS generation mechanism.   Fixing the initial conditions, 
a well defined qualitative behavior is observed in all the simulations. We usually set 
$F_l$ (that is, the nutrient and the type of bacteria) and vary  the flow or  the 
concentration. The parameter values we use do not correspond to any specific 
bacteria-nutrient choice, since adequate experimental data for parameter calibration 
are not available yet.


The first numerical experiments show three dimensional patterns with
$\sigma({\cal C})=\sigma$, and $\tau(Re)$ also a constant. 
The ratio ${\tau(Re) \over \sigma}$ 
is a control parameter. The initial seed is the same in all the tests: a slab
containing $110 \times 20 \times 4$ cells. Two erosion processes are observed: 
sloughing of large fragments, and smooth erosion of surfaces.
When the shear is large enough compared to the biofilm cohesion 
(${\tau \over \sigma}$ large) and the limiting concentration is not too high, 
the initial layer of biofilm remains almost flat and homogeneous while it is
slowly washed out, see Figure \ref{washout}. Cells are eroded from the 
front, but may grow downstream.
Biofilms that are not strong enough  have been experimentally observed to 
be washed out on  glass surfaces in \cite{stoodleycohesion}. This happens
typically to biofilms created at lower Reynolds number when the shear is
increased.

Decreasing the ratio ${\tau \over \sigma}$ or increasing slightly the concentration, 
the initial biofilm seeds develop ripple-like patterns, that advance downstream 
with the flow, as in Figure \ref{ripple}. 
Ripples anchor and become streamer-like structures (fingers or peaks elongated 
in the direction of the flow) for smaller ${\tau \over \sigma}$ or larger concentrations, 
see Figures \ref{streamer} and \ref{variant}(a). Fingers that become too large 
may detach.
Ripples traveling downstream on top of lower layers of biofilm have been 
reported for both laminar and turbulent flows in \cite{stoodleyfigures,stoodleyripples}. 
Networks of streamers being eroded and leaving small ripples behind
have been experimentally observed in \cite{stoodleyfigures}. 
Low enough ratios ${\tau \over \sigma}$ or large enough concentrations
lead to networks of mounds or towers separated by voids, as in Figure
\ref{variant}(b) and Figure \ref{mound}, where $F_l$ is increased so that towers
are more clearly noticed with fewer cells. Similar patterns are commonly 
observed in nature in low shear environments, see \cite{costerton2}. 

The evolution of the cellular aggregates reproduced in these 3D simulations
includes only growth and erosion processes. What are the precise 
mechanisms producing different ripples or streamers observed in real 
biofilms (in both laminar and turbulent regimes) is uncertain, though 
mechanical effects are believed to be relevant. Cell displacement due
to the flow, decay and cell adhesion should also be considered to gain 
insight on the processes that trigger real pattern formation. 

\begin{figure}[!ht]
\begin{center}
\includegraphics[width=15cm]{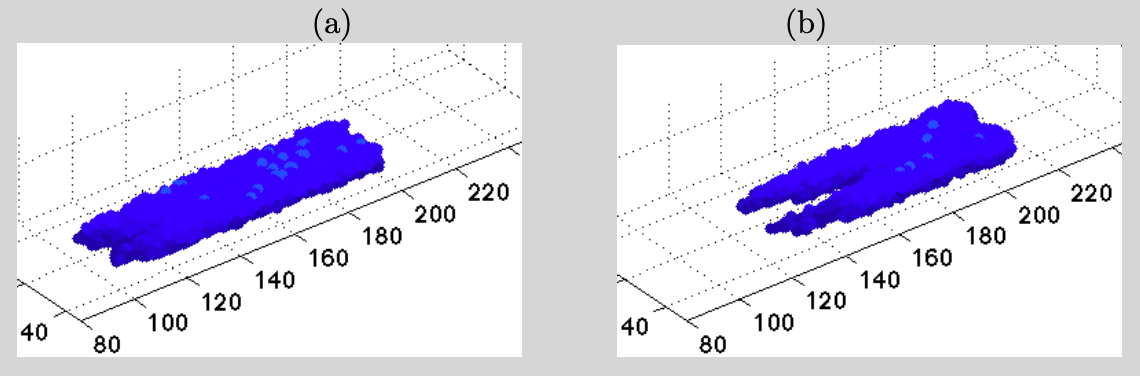} 
\end{center}
\caption{
An initially flat and homogeneous biofilm being washed out by the 
current. Snapshots are taken at steps $T=20$ and $T=60$. After $200$ 
steps all cells have eroded. Dimensionless parameter values: 
$\hat{C}_l=2.25$, $F_l=0.04$,  $\delta_B=5$, ${\tau(Re) \over \sigma}=5$, $\beta=1$. }
\label{washout}
\end{figure}

\begin{figure}[!ht]
\begin{center}
\includegraphics[width=15cm]{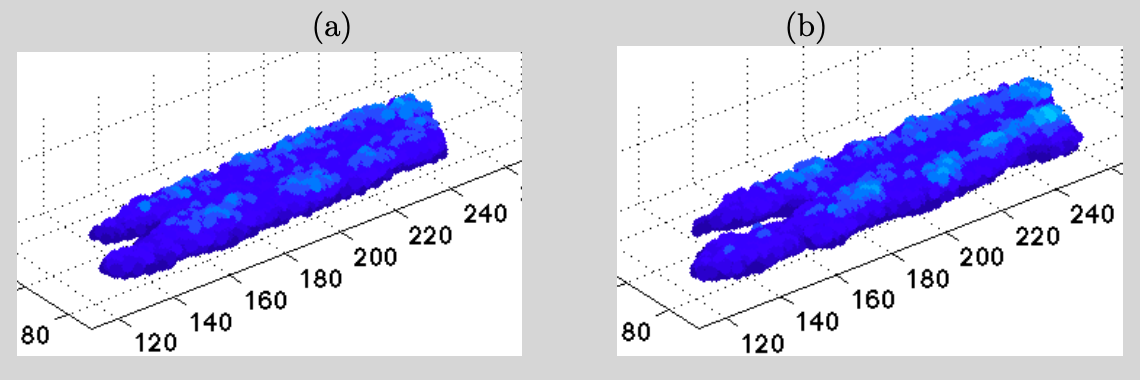} 
\end{center}
\caption{ 
A  flat and homogeneous biofilm seed develops ripple-like
patterns moving downstream with the flow. Snapshots are taken 
at steps $T=50$ and $T=80$. A peak located at grid position $200$ 
clearly migrates to position $220$. Same parameter values as in 
Fig. \ref{washout} except ${\tau(Re)\over \sigma}=2$.}
\label{ripple}
\end{figure}

\begin{figure}[!ht]
\begin{center}
\includegraphics[width=15cm]{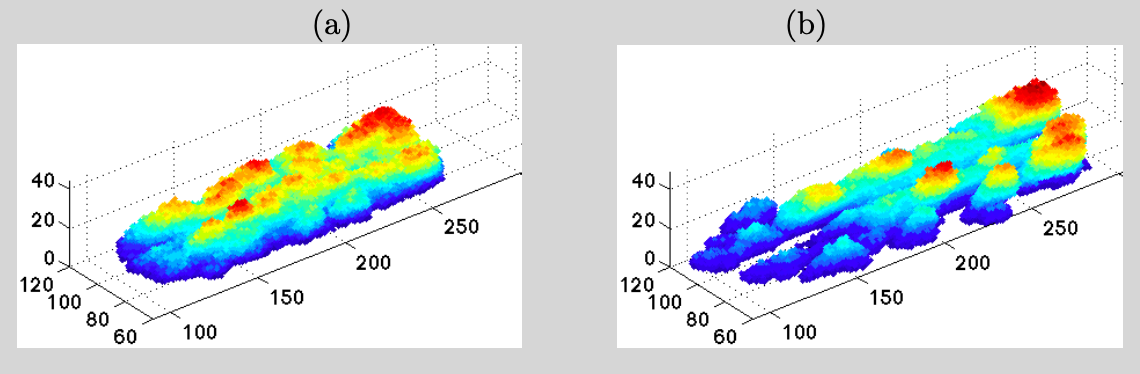} 
\end{center}
\caption{A flat and homogeneous biofilm seed generates
streamer-like structures.
Same parameter values as in Fig. \ref{ripple}  except
${\tau(Re)\over \sigma}=1.5$. Snapshots are taken at steps 
$T=80$ and $T=160$ ($75089$ alive cells).}
\label{streamer}
\end{figure}

\begin{figure}[!ht]
\begin{center}
\includegraphics[width=15cm]{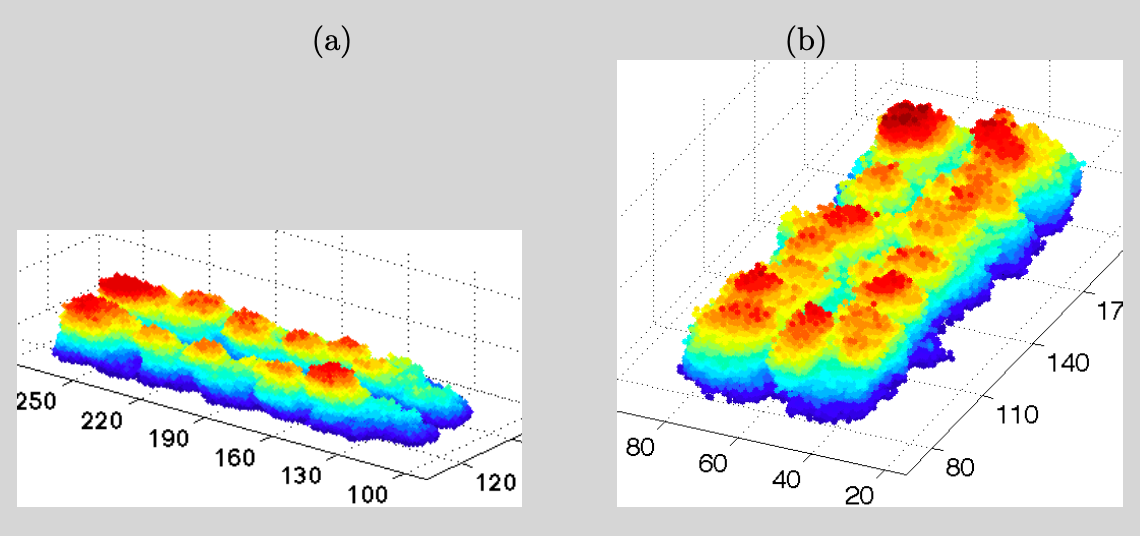} 
\end{center}
\caption{ (a) Increasing the outer concentration in Fig. \ref{ripple} we also 
find streamer-like structures. Snapshot taken at $T=80$ for $\hat{C}=3$
($71516$ alive cells).
(b) Further decreasing ${\tau(Re)\over \sigma}$ in Fig. \ref{streamer}
we generate mounds. Snapshot taken at $T=70$  for 
${\tau(Re)\over \sigma}=0.5$ ($88815$  alive cells).}
\label{variant}
\end{figure}

\begin{figure}[!ht]
\begin{center}
\includegraphics[width=15cm]{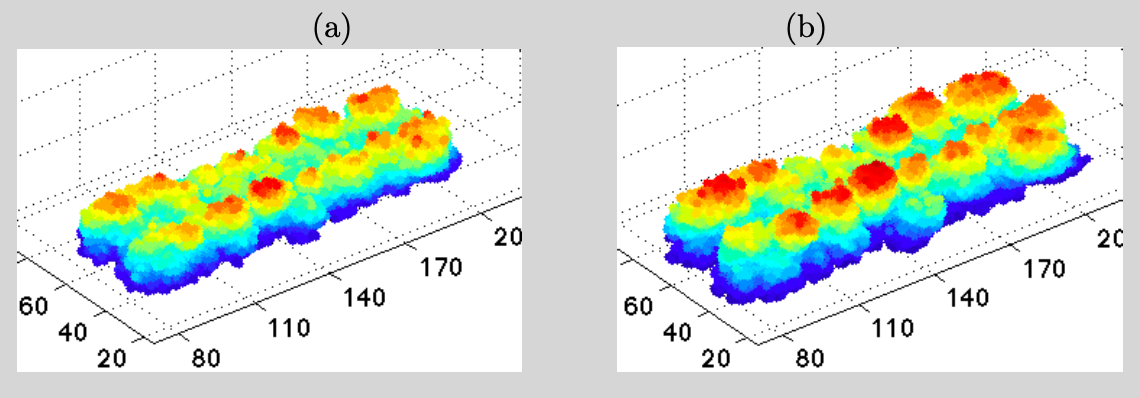} 
\end{center}
\caption{A biofilm seed forms a group of mounds.
Same parameter values as in Fig. \ref{variant}(b) except
$F=0.08$. Snapshots are taken at steps 
$T=70$ and $T=100$ ($63534$ alive cells).}
\label{mound}
\end{figure}

The observed patterns depend on the combination of parameters
selected: $F_l$, $\delta_B$, $\hat{C}_l$ and ${\tau \over \sigma}$. 
When only growth processes are taken into account, vertical fingers
are formed if the growth dynamics is constrained
by the limiting concentration, see \cite{fingers,poplawski}. If a large 
enough concentration reaches most cells, biofilms tend to be flatter. 
According to the equations that govern the concentration, 
different factors may favor this:
\begin{itemize}
\item Decreasing $F_l$ reduces nutrient/oxygen uptake by the cells
and increases the average values of the limiting concentration field
for a fixed biofilm geometry, $\delta_B$ and $\hat{C}_l$. $F_l$
depends on the nutrient type and the bacterial species through
uptake rates, saturation and diffusion coefficients. Therefore,
it is fixed for specific choices of bacteria and nutrient. 
The threshold concentration to hinder fingering can only be reached 
if the limiting concentration in the fluid $\hat{C_l}$ is large enough 
depending on $F_l$ and the biofilm thickness.
\item Decreasing the boundary layer thickness $\delta_B$ also 
increases the concentration field, for a fixed biofilm geometry, $F_l$ and 
$\hat{C}_l$. However,  concentrations that are large enough uniformly  can 
only be reached if the limiting concentration in the fluid $\hat{C}_l$ is large 
enough. In practice, $\delta_B$ decreases with the flow velocity. Rising the 
flow velocity might increase everywhere the concentration field and favor 
flatter biofilms.
\item Choosing $\hat{C}_l$ large enough depending on the biofilm geometry, 
$F_l$ and $\delta_B$, we may end up with flatter biofilms. Their 
thickness will increase with time, unless we are able to choose a  current 
strong enough to keep them thin. However, that might in turn generate patterns
by a different mechanism, or just wash out the biofilm on smooth
surfaces.
\end{itemize}
In intermediate regimes we may see all sorts of dendritic or porous
patterns,  especially in two dimensions. The above remarks apply in static
or very slow flows. 
As the ratio ${\tau \over \sigma}$  grows, the erosion mechanism modifies the 
picture. It may modify or suppress patterns or create additional ones depending 
on the values of ${\tau \over \sigma}$ and $\delta_B$. Strong erosion does not 
require large flows, small biofilm cohesion suffices.
Increasing the concentration, we may reduce or overcome erosion effects
as in Figures \ref{variant}(a) or \ref{fig89}(b).

Two dimensional simulations produce similar results to 3D simulations, 
though the patterns tend to be more branchy. The thresholds separating flat 
biofilms, wavy biofilms, mounds and streamer-like or mushroom-like patterns 
are shifted. Depending on $F_l$ and $\hat{C}_l$, some of the regimes observed 
as we vary ${\tau \over \sigma}$ may almost vanish. 

We revisit our three dimensional simulations on smooth and rough surfaces 
using the two dimensional model. Almost all the simulations start from an
initial biofilm seed containing $110 \times 4$ cells.
Roughness adds another variable that interacts with the growth and 
erosion mechanisms. 
We represent it  by the peaks depicted in Figure \ref{fig24}. 
Some roughness patterns may prevent the washing out effect described
before as ${\tau \over \sigma}$ increases. In Figure \ref{fig81} (c)-(d), 
$2 \times 2$ square peaks with an interpeak space of $5$ tiles help the
biofilm the remain attached to surfaces they have already colonized and expand 
onto neighboring downstream  regions. It is washed out on a smooth surface,
see Figure \ref{fig81} (a)-(b).
Reducing  the spacing between peaks to $2$ tiles, we see a similar behavior with 
less cells and flatter biofilms, since cells have less space to reproduce.  
Increasing the spacing to $10$ tiles, the evolution is similar. However, biofilms 
contain more biomass since cells  have more space to divide and spread.  
If this spacing increases further, the sheltering effect of roughness decreases. 
If the depth of the steps increases, nutrients become too scarce to sustain the 
cell population, hindering biofilm survival.
Considering several biofilm peaks as initial data instead of flat layers, we
see a similar evolution. Biofilms are washed out on a flat substratum. 
Surface roughness of the same order of magnitude as the bacterial size helps
the peaks to colonize neighbouring regions and merge, producing
more uniform covers, as in Figure \ref{fig83}. Biofilms succeed in expanding 
upstream. Other roughnesses may have different effects.

Decreasing the ratio ${\tau(Re) \over \sigma}$, we find a new regime in which 
ripple-like patterns are observed. 
Newborn cells accumulate downstream, forming ripples and a floating finger 
at the rear of the biofilm, which eventually reattaches allowing the biofilm to expand 
downstream, see Figure \ref{fig85} (a)-(b).  More limited upstream expansion
may also occur. Roughness anchors the front of the biofilm, fostering colonization 
of downstream  regions, as in Figure \ref{fig85} (c)-(d).  Thicker biofilms with 
more noticeable ripples seem to be formed.
Further decreasing ${\tau(Re) \over \sigma}$, fingers aligned with the flow develop, 
which detach when they surpass a certain size.    
Figures \ref{fig87} (a)-(b) show  the biofilm evolution on a flat substratum.  
For smaller ${\tau(Re) \over \sigma}$, branchy biofilm towers separated by 
fluid grow on the initial biofilm, as in  Figure \ref{fig89} (a). 
When $\hat{C}_l$ is increased, mushrooms are more easily formed, that 
become denser and merge for higher values of $\hat{C}_l$, as in Figure 
\ref{fig89} (b).

\begin{figure}[!t]
\centering
\includegraphics[width=15cm]{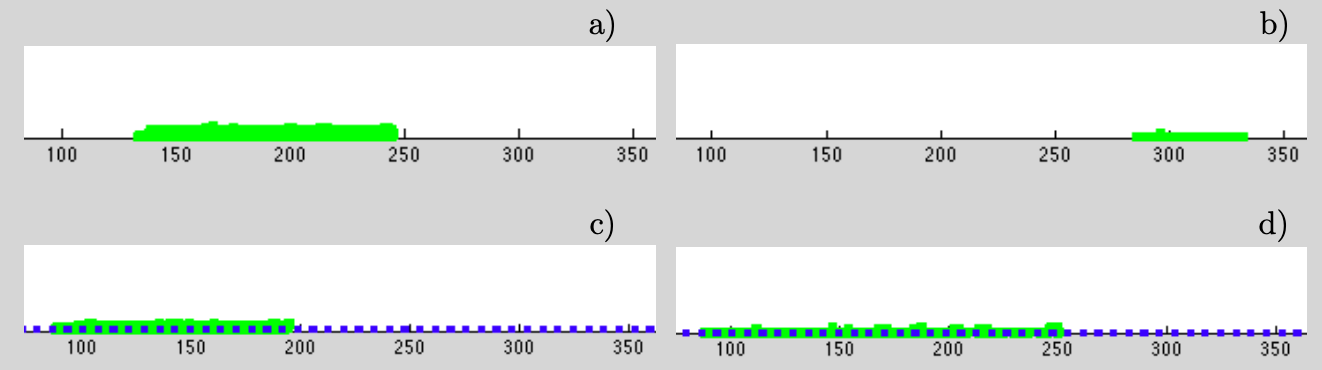}
\caption{
(a), (b) Snapshots showing a biofilm being washed out  over a flat substratum for large enough ${\tau(Re) \over \sigma}$.
(c), (d) On a rugose surface, with steps characterized by peak height $\epsilon=2$, length $\lambda=2$, and interpeak distance $\delta=5$, the  biofilm front is anchored and the biofilm expands colonizing new regions downstream.  
Dimensionless parameter values: ${\tau(Re) \over \sigma} =5$,  $F_l=0.04$, $\hat{C}_l=1.5$, $\delta_B=5$, $\beta=1$. Time between snapshots: $450$ time steps.}
\label{fig81}
\end{figure}

\begin{figure}[!t]
\centering
\includegraphics[width=15cm]{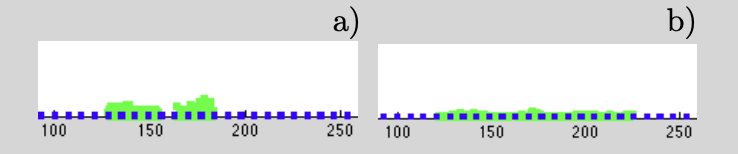} 
\caption{Same as Figure \ref{fig81} starting from two small colonies, which merge and expand helped by roughness, downstream but also slightly upstream. On a flat substratum they are washed out.}
\label{fig83}
\end{figure}

\begin{figure}[!t]
\centering
\includegraphics[width=7.5cm]{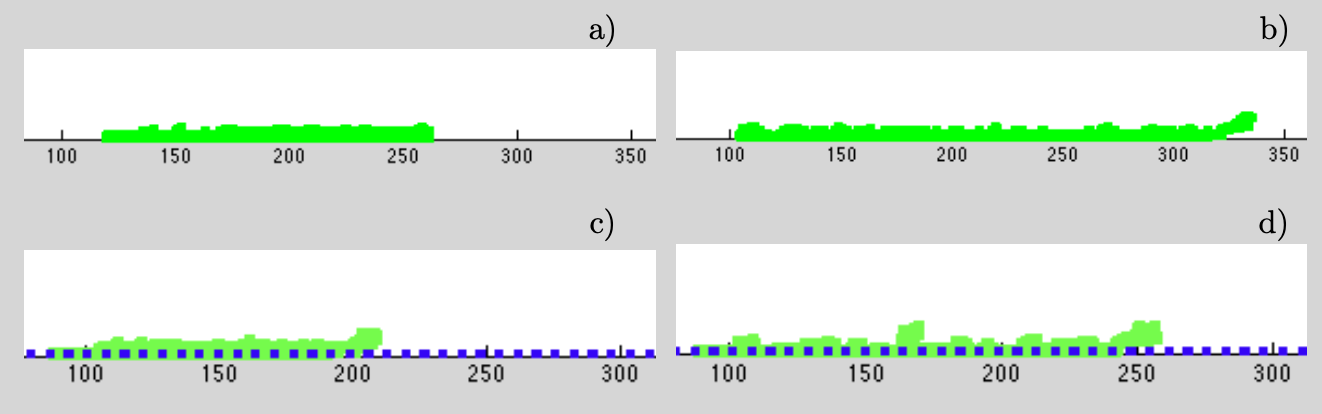} 
\caption{Ripple-like structures form on an initially flat seed: (a), (b)  over a flat substratum, 
(c), (d) on a rugose surface with steps characterized by peak height $\epsilon=2$, length $\lambda=2$, and interpeak distance $\delta=5$. Dimensionless parameter values: ${\tau(Re)\over \sigma}=2$,  $F_l=0.04$, $\hat{C}_l=1.5$, $\delta_B=5$, $\beta=1$. Time between snapshots: $800$ time steps.}
\label{fig85}
\end{figure}

\begin{figure}[!t]
\centering
\includegraphics[width=15cm]{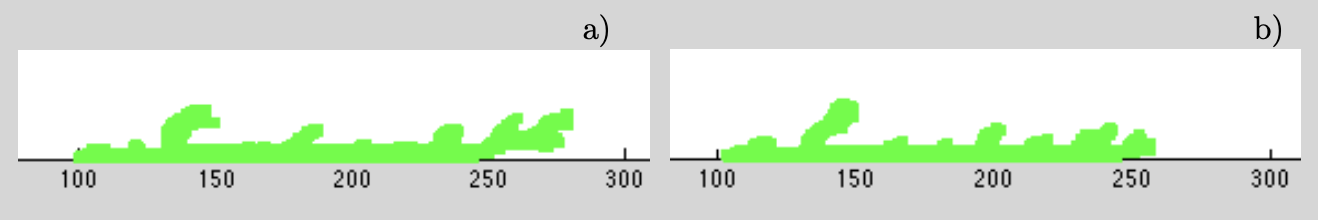} 
\caption{
Fingers curved with the flow develop in a biofilm over a flat substratum:
 a) Time step $T=300$. (b) Time step $T=500$. The trailing finger has
eroded, it will grow  again later. The biggest finger at the front detached
but has grown again. An intermediate finger has also detached, leaving
room for another one at its back to grow.
Dimensionless parameter values: ${\tau(Re)\over \sigma}=1$, $F_l=0.04$, 
$\hat{C}_l=1.5$, $\delta_B=5$, $\beta=1$.   }
\label{fig87}
\end{figure}

\begin{figure}[!t]
\centering
\includegraphics[width=15cm]{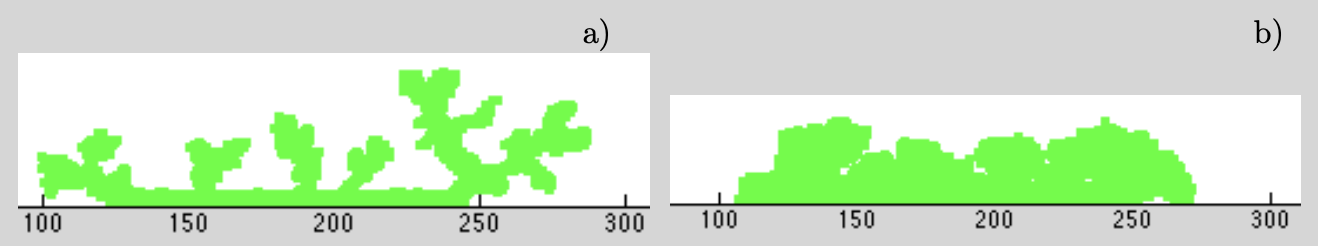}  
\caption{(a) Same as Figure \ref{fig87} for smaller ${\tau(Re)\over \sigma}=0.25$. An 
expanding  network of biofilm branches infiltrated with fluid is  formed.
(b) Same as Figure \ref{fig87} increasing the limiting concentration to
$\hat{C}_l=2.7$. The biofilm expands forming dense mushrooms separated by narrow 
channels, that eventually merge.}
\label{fig89}
\end{figure}

All the previous numerical simulations set $\beta=1$ in (\ref{shearmod}). 
That assumes that cells only feel the influence of the flow in the $x$ direction. 
We have repeated the previous simulations for $\beta \in (0.9,0.99)$ finding 
an increasing chance of  long biofilm layers splitting in smaller patches, that 
may eventually be washed out, and a reduction in the maximum height of 
the patterns.  Otherwise, similar trends are observed provided the concentration
is large enough to compensate the increased erosion.
In Figure \ref{beta} (a), an initially flat biofilm seed breaks into separated patches.  
Increasing the concentration,  the biofilm splits in patches that evolve into 
streamer-like patterns, see Figure  \ref{beta} (c). Decreasing ${\tau \over \sigma}$, 
a stable network of biofilm towers separated by fluid channels develops in Figure  
\ref{beta} (d). 
Figure \ref{beta} (b) shows the effect of roughness, compared to subplot (a).
In practice, as $Re$ grows turbulent effects may be important and the model 
should allow for a degree of erosion in the $z$ direction. As $Re$ increases, 
$\beta(Re)$ should probably diminish, enhancing erosion by the flow.
However, there are a number of overlapping competing effects with uncertain
outcome. The thickness of the concentration boundary layer $\delta_B(Re)$ 
should also decrease,  augmenting the concentration of nutrients, and growth 
rates thereof. In case floating cells are present, adhesion rates might equally 
increase with $Re$, leading to larger biofilm accumulation \cite{stoodleysulfate}. 

\begin{figure}[!ht]
\centering
\includegraphics[width=15cm]{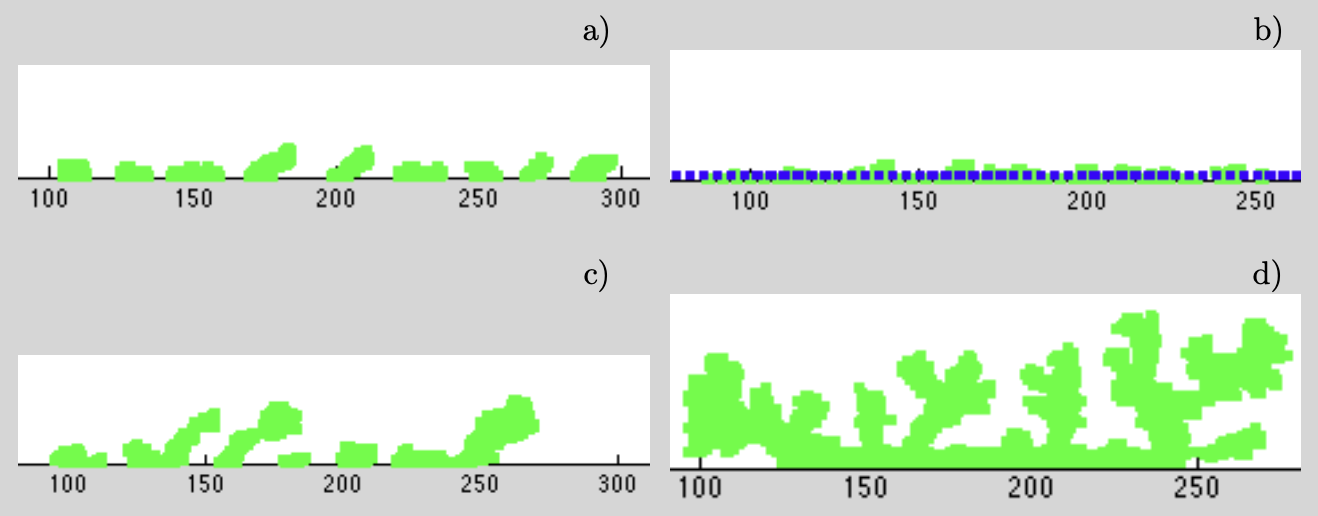} 
\caption{Evolution of a flat biofilm seed when $\beta\neq 1$.
(a) ${\tau \over \sigma}=2$, $\hat{C}_l=1.5$, $\beta=0.99$, 
$\delta_B=5$, $F_l=0.04$, $T=600$.
The biofilm breaks into patches that generate small peaks.
(b) Same parameters as (a) except  ${\tau(Re) \over \sigma}=5$, 
on a rough surface. Fingers are eventually eroded, leaving a thin 
biofilm behind.
(c) Same parameters as (a) with increased $\hat{C}_l=1.8$. Streamer-like
structures are formed, that detach when they become too large.
(d) Same parameters as (c) with  smaller ${\tau(Re) \over \sigma}=0.25$. 
An expanding network of biofilm branches develops.
}
\label{beta}
\end{figure}

So far we have only considered the evolution of attached biofilm 
seeds, neglecting the possible adhesion of floating cells. 
Figure \ref{adhesion1} shows the evolution of
an uncolonized substratum under a flow carrying
nutrients and bacteria. The initial state is the same in all the trials, 
a clean surface. The density of 
bacteria, oxygen and nutrients carried by the flow is also the
same in all the figures.  The ratio ${\tau \over \sigma}$ and the
number of attached cells $N$ vary.  
For high enough adhesion rates the biofilm may cover completely the 
bottom of the pipe provided the rate of adhesion of   bacteria 
is high enough compared to the erosion effects, see Figure
\ref{adhesion1}(a). Biofilm patches form, which eventually merge. 
After some time, the substratum is 
fully covered by wavy biofilm layers of increasing thickness. 
Otherwise, only patches, or isolated  peaks grow, see Figure 
\ref{adhesion1}(b). 
It has been experimentally observed in \cite{unpublished} that low
adhesion rates at low Reynolds numbers lead to patchy configurations,
see Figure \ref{fig4}. Larger adhesion rates at larger Reynolds
numbers have produced rippled biofilm layers, see Figure \ref{fig5}.

\begin{figure}[!ht]
\centering
\includegraphics[width=15cm]{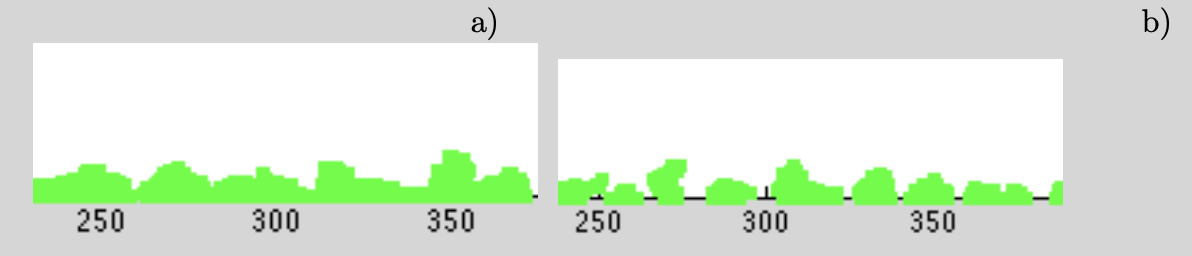} 
\caption{(a) Biofilm growth at large adhesion rates. A wavy biofilm
carpet is formed when ${\tau(Re) \over \sigma}=2.5$ and $N=80$.
(b) Biofilm grown at smaller adhesion rates. Scattered patches are
generated when ${\tau(Re) \over \sigma}=1$ and $N=10$.
Other dimensionless parameter values are $\hat{C}_l=1.5$, $\beta=0.99$, 
$F_l=0.04$ and $\delta_B=5$. Snapshots are taken at  time step $75$. 
}
\label{adhesion1}
\end{figure}

The length scales in the patterns when erosion effects are low seem to be governed by the average concentration the cells feel. Low concentrations seem to increase the distance between patterns. Raising $F_l$ and $\delta_B$ produces that effect.
Figures \ref{fig814}(a)-(b) show the effect of $\delta_B$. For small $\delta_B$, flat dense biofilms may be formed, see Figure \ref{fig814}(a). Small $F_l$ would have a similar effect. Increasing $F_l$ or decreasing $\hat{C}_l$, protuberances appear again. For larger $\delta_B$, fingers are formed at increasing distances, see Figures \ref{fig87}, \ref{fig89}, \ref{fig814}(b), or Figure \ref{adhesion4}(a), which includes also adhesion.  Figures \ref{variant} (b) and \ref{mound} illustrate variations in the patterns with $F_l$.
Taking only into account erosion and growth, the boundary layer thickness $\delta_B$  was shown in \cite{hermanovic} to control the distance between patterns for fixed $F_l$.  Notice that in \cite{hermanovic}, $\tau({\cal C})$ in (\ref{shearmod}) is set equal to a constant and the resulting fingers do not see the direction of the flow.  In static flows, the distance between fingers is shown to increase as the availability of nutrients diminishes in \cite{poplawski}.  
Additionally, the erosion mechanism can generate, suppress or modify patterns. It can erode and curve the fingers that would grow in absence of flow or change initial distances between fingers. Some of them may detach when they surpass a certain size, leaving just a few, or only one at the end, or they may detach and grow in turn. Adhesion mechanisms further modify the picture, see Figure \ref{adhesion4} (b), where peaks have been replaced by wavy layers.
 
\begin{figure}[!t]
\centering
\includegraphics[width=15cm]{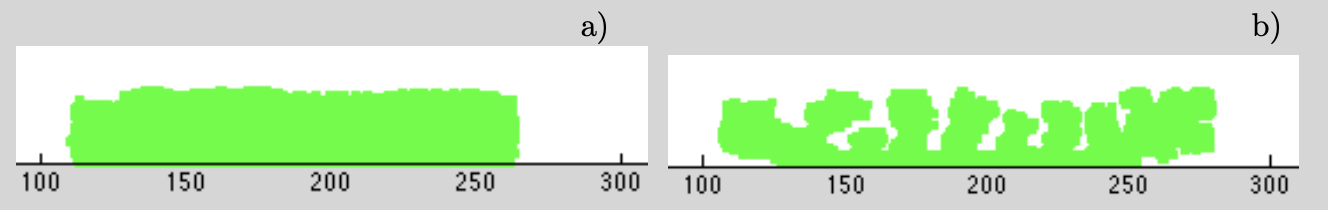} 
\caption{(a) Biofilm for a small boundary layer thickness $\delta_B=1$. A flat 
biofilm seed evolves into a thickening dense biofilm.
(b) Biofilm evolution for an increased boundary layer thickness $\delta_B=3$.
A mushroom network develops.
Other dimensionless parameter values are ${\tau(Re) \over \sigma_0}=2$,
$\hat{C}_l=1.5$, $\beta=1$,  and $F_l=0.04$. Snapshots are taken at  time 
steps $25$ and $75$, respectively.}
\label{fig814}
\end{figure}

\begin{figure}[!ht]
\centering
\includegraphics[width=15cm]{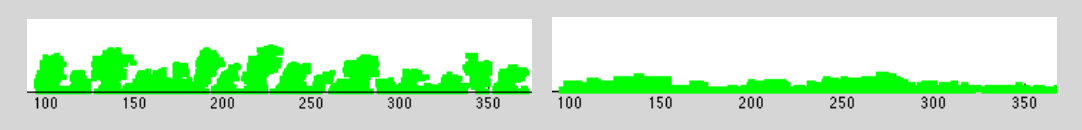}
\caption{ Variations of the boundary layer thickness including adhesion.
(a) Biofilm evolution after $50$ time steps when  $\delta_B=3$. A network
of biofilm towers separated by fluid  is grown.
(b) Biofilm evolution  after $50$ time steps when $\delta_B=7$.
The surface is covered by a wavy carpet of biofilm.
Other parameter values are are $\hat{C}_l=1.5$, $\beta=0.99$, 
${\tau \over \sigma}=2$, $N=60$, $F_l=0.04$.
}
\label{adhesion4}
\end{figure}

In the previous figures, the cohesion parameter $\sigma$ was set equal to a constant. 
The remaining figures incorporate the EPS generation mechanism for $\sigma$ given by   
(\ref{sigma}) in absence of floating cells.
Compare Figures \ref{fig85} and \ref{fig87} to Figures \ref{fig810}(a)-(b).
Resulting biofilms are more rigid.
As ${\tau(Re)\over \sigma_0(Re)}$ is further reduced, biofilm towers are stronger and 
do not deviate in the direction of the flow, see Figure \ref{fig810}(c).
In these figures, $\sigma$ varies locally, depending on the number of bacteria generating
EPS matrix, and the values we give to the parameters $\sigma_0$ and $\alpha$,
representing the strength of the EPS matrix and standard attached cells of
a specific bacteria species.   The percentage of cells producing EPS matrix remains stable 
during the biofilm evolution and might be used to calibrate $\sigma_0$ and $\alpha$,
since this is a parameter that can be measured experimentally in real biofilms.
Matrix generation affects growth and local consistency. The thresholds separating 
different pattern regimes are apparently shifted. 

\begin{figure}[!t]
\centering
\includegraphics[width=7.5cm]{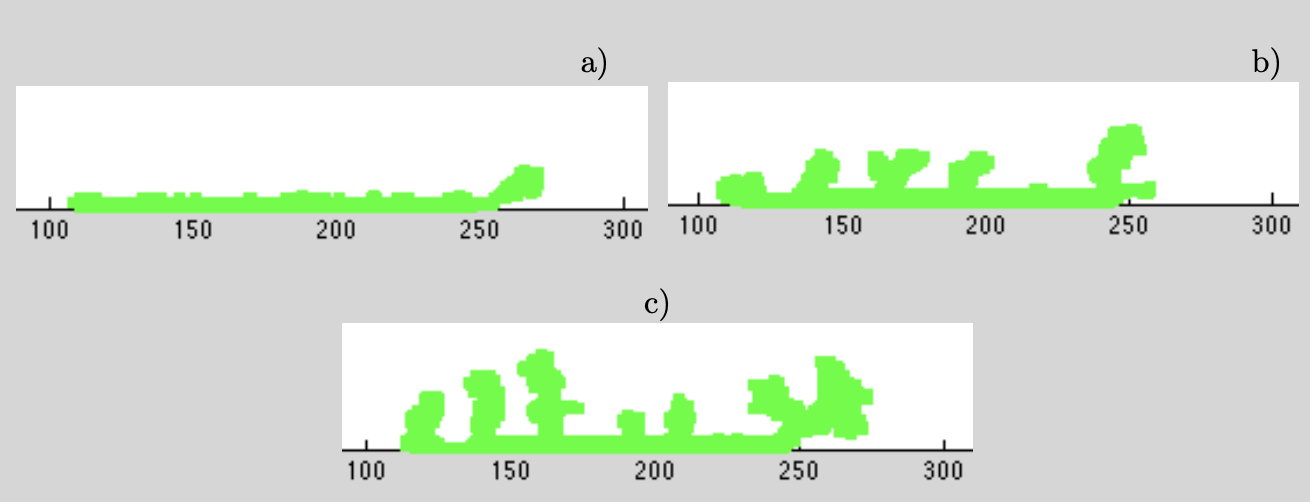}
\caption{Tests including the EPS generation mechanism. The constant cohesion parameter $\sigma$ is replaced by a self-adjusted variable cohesion parameter 
$\sigma$ given by (\ref{sigma}): 
(a) Small ripples are formed when ${\tau(Re)\over \sigma_0(Re)}=2$, $R(Re)=0.4$.
(b) Peaks curved in the direction of the flow develop on more compact biofilms when ${\tau(Re)\over \sigma_0(Re)}=0.5$. 
(c) Vertical towers appear when ${\tau(Re)\over \sigma_0(Re)}=0.25$. The symmetry of the pattern shows a decreasing dependence on the flow direction.
Other dimensionless parameter values: $\hat{C}_l=1.5$, $\delta_B=5$, $\beta=1$, $F_l=0.04$, $T=150$.
}
\label{fig810}
\end{figure}

In general, the role of the controlling parameters once the type of bacteria and nutrient is fixed seems to be the following. 
In absence of floating cells, the parameter $F_l$ appears to regulate the critical thickness for the biofilm to survive and the regimes for patterns. Small $F_l$ due to small uptake rates may result in mostly flat patterns. Large $F_l$ due to small diffusivities may result in mostly tower-like structures. 
Once $F_l$ is fixed, $\delta_B$ has a similar effect and the ratio between the shear due to the flow and the biofilm cohesion determines the degree of erosion of the biofilm by the flow. It competes with the concentration to determine the patterns. 
For low concentrations, we switch from vertical to curved fingers, then ripples and finally homogeneous biofilms as the ratio of the shear to the biofilm cohesion increases. As the concentration grows, the transition between these regimes occurs at larger ratios. We usually end up with a mixture of bacterial towers infiltrated with networks of channels, unless the concentration is too large and bacteria fill all the available space.
Adhesion of floating cells may change the pattern regimes. Low adhesion rates may produce patchy biofilms whereas large adhesion rates may generate wavy biofilm layers, depending on erosion
and growth. One must keep in mind that biofilm evolution depends
on how many mechanisms are relevant in the time scale we are woking in, and
what all the values of the main controlling parameters.


\section{Conclusions}
\label{experiments}

A  stochastic model for submerged biofilm growth on rugose surfaces is proposed. Cell behavior 
is governed by a set of probabilistic rules for cell division, spreading, EPS generation,  cell 
detachment,{ cell deactivation and adhesion}.  Different patterns are generated as a result of the 
collective behavior of cells acting according to the rules.
Insight on the interplay of the competing mechanisms considered is gained. { 
Combinations of mechanisms producing patterns similar to structures observed
in real biofilms are identified.}
The proposed framework for studying the behavior of cell aggregates is quite flexible, and 
may be used to test mechanisms for cell behavior or be combined with more refined descriptions 
of biofilm evolution.
The parametric study we have performed reproduces some qualitative trends already  observed 
in other groups experiments \cite{paureaginosa,stoodleyfigures,stoodleynutrients,stoodleyripples, stoodleysulfate,stoodleycohesion} and in our own experiments. 
{ Our model does not account yet for mechanical processes like
cell displacement within the biofilm or movement of biofilm blocks due to external forces,
which are thought to be relevant in the formation of real biofilm  streamers. 
Further work on this issue is needed to assess its effect on the observed patterns.}

{ In absence of floating cells, the evolution of a biofilm seed} depends
on a main set of parameters: the ratio of uptake rates to diffusional supply, the ratio of 
the shear due to the flow to the biofilm cohesion, the thickness of the concentration 
boundary layer and  the values of the concentrations in the outer fluid. {  Erosion and 
growth mechanisms alone are able to generate biofilm structures moving downstream.}
When the nutrient type and the bacterial species are fixed, different patterns are
generated as the shear due to the flow or the concentration of oxygen and nutrients
inside the flow vary: networks of ramified towers separated by fluid channels, vertical
fingers, streamer-like structures, ripple-like patterns traveling downstream,  
flat  biofilms, and so on.  For slow or static flows, fingering is avoided when
the limiting concentration reaches easily all biofilm cells. Erosion affects the growth
of fingers: they may deviate in the direction of the flow or their heights
may be severely reduced. Thickening flat biofilms may be eroded and kept
thin. Strong erosion does not require large flows. Small biofilm cohesion is enough.
For most nutrients, increasing the ratio of the shear due to the flow to the biofilm
cohesion we find more homogeneous  and thinner biofilms, that may be eventually 
washed out, depending on the concentration.
Surface roughness of the same order of magnitude as the bacterial size may anchor
the biofilm and promote its survival and expansion.
Other types of roughness may hinder biofilm growth.
{ Adhesion of floating cells can change qualitatively the nature of the observed patterns.  
For small adhesion rates, patchy biofilms may be formed on clean surfaces. For larger 
adhesion rates, wavy uniform covers may appear, depending on the remaining
parameters.}  


The  insight gained  on the influence of different variables in the evolution of biofilms 
may be useful to control their structure, either
to destroy them or to use them to our advantage in different technological,
medical and environmental problems.

\acknowledgments
The authors thank V. de Lorenzo, E. Mart\'{\i}nez (Centro Nacional de Biotecnolog\'{\i}a, Spain) and A. Vel\'azquez, J.R. Arias (Universidad Polit\'ecnica de Madrid, Spain) for experimental support and fruitful discussions. 
D. Rodr\'{\i}guez and A. Carpio were supported by the Autonomous Region of Madrid and the spanish Ministry of Research through grants S2009/DPI-1572, and FIS2011-28838-C02-02, FIS2010-22438-E. A. Carpio was also supported by a mobility grant of Fundaci\'on Caja Madrid. B. Einarsson was supported by a grant of the NILS program and project FIS2008-04921-C02-01. A. Carpio and D. Rodriguez thank
M.P. Brenner for hospitality at Harvard University.


\begin{thebibliography}{9}
\bibitem{nota}   The size $a$ of one bacterium   is of order $1\mu \,m$. In square channels whose section is of order $b=1 \, mm$-$100 \mu \,m$ and have steps of size $a$, $a$ itself is a reasonable choice as spatial step to discretize the reaction diffusion-equations governing the concentrations or the Navier-Stokes equations for the external flow.
\end{thebibliography}
\end{document}